\documentclass[journal,twoside,web]{ieeecolor}
\usepackage{tmi}
\usepackage{cite}
\usepackage{diagbox}
\usepackage[misc]{ifsym}
\usepackage{amsmath,amssymb,amsfonts}
\usepackage{algorithmic}
\usepackage{subfigure}
\usepackage{graphicx,subfigure}
\usepackage{textcomp}
\usepackage{booktabs}
\usepackage[OT1]{fontenc} 

\def\BibTeX{{\rm B\kern-.05em{\sc i\kern-.025em b}\kern-.08em
    T\kern-.1667em\lower.7ex\hbox{E}\kern-.125emX}}
\begin{document}
\title{ aiWave: Volumetric Image Compression with 3-D Trained Affine Wavelet-like Transform}
\author{Dongmei Xue, Haichuan Ma, Li Li, \IEEEmembership{Member, IEEE}, Dong Liu, \IEEEmembership{Senior Member, IEEE}, Zhiwei Xiong,
\IEEEmembership{Member, IEEE}
\thanks{
This work was supported in part by the Natural Science Foundation of China under Grant 62171429 and 62021001; in part by the Fundamental Research Funds for the Central Universities under Grant WK3490000006; This work was also supported by USTC Research Funds of the Double First-Class Initiative Grant YD3490002001. It was also supported by the GPU cluster built by MCC Lab of Information Science and Technology Institution, USTC.}
\thanks{The authors are with the CAS Key Laboratory of Technology in Geo-Spatial Information Processing and Application System, University of Science and Technology of China, Hefei 230027, China. Zhiwei Xiong is also with the Institute of Artificial Intelligence, Hefei Comprehensive National Science Center, Hefei 230088, China (E-mail: xdm1@mail.ustc.edu.cn; hcma@mail.ustc.edu.cn; lil1@ustc.edu.cn; dongeliu@ustc.edu.cn; zwxiong@ustc.edu.cn). Dr. Li Li is the corresponding author.}}

\maketitle

\begin{abstract}
Volumetric image compression has become an urgent task to effectively transmit and store images produced in biological research and clinical practice. 
At present, the most commonly used volumetric image compression methods are based on wavelet transform, such as JP3D.
However, JP3D employs an ideal, separable, global, and fixed wavelet basis to convert input images from pixel domain to frequency domain, which seriously limits its performance.
In this paper, we first design a 3-D trained wavelet-like transform to enable signal-dependent and non-separable transform.
Then, an affine wavelet basis is introduced to capture the various local correlations in different regions of volumetric images. 
Furthermore, we embed the proposed wavelet-like transform to an end-to-end compression framework called aiWave to enable an adaptive compression scheme for various datasets.
Last but not least, we introduce the weight sharing strategies of the affine wavelet-like transform according to the volumetric data characteristics in the axial direction to reduce the number of parameters.
The experimental results show that:
1) when cooperating our trained 3-D affine wavelet-like transform with a simple factorized entropy coding module, aiWave performs better than JP3D and is comparable in terms of encoding and decoding complexities; 
2) when adding a context module to remove signal redundancy further, aiWave can achieve a much better performance than HEVC.  

\end{abstract}
\begin{IEEEkeywords}
Volumetric image compression, wavelet-like transform, lifting scheme, convolutional neural network, end-to-end image compression
\end{IEEEkeywords}

\section{Introduction}
\label{sec:introduction}

\IEEEPARstart{V}{olumetric} images, including biological and medical images, are widely utilized nowadays due to the advance in imaging technology.
Among all kinds of volumetric images, the biological images are mainly obtained by electron microscopy (EM) and play an essential role in investigating the morphology and pathology of tissue cells. 
Medical images such as computer tomography (CT) and magnetic resonance imaging (MRI) have been applied to clinical diagnosis in most modern hospitals.
However, the massive size of volumetric images brings considerable challenges to hardware storage and transmission.
For example, the raw brain images of drosophila occupy approximately \(106\) terabytes (TBs) of storage space\cite{zheng2018complete}.
To this end, exploring effective volumetric image compression techniques has become an urgent task.
In recent years, many methods have been proposed to address this challenge\cite{wang2017learning,lucas2017lossless,rossinelli2020high,hernandez2018mosaic,gao2020volumetric,gao2019deep}. 
They can be roughly divided into three groups: wavelet-based methods, video-based methods, and autoencoder-based methods.

The most commonly used methods for volumetric image compression are based on wavelets. 
One representative method is JP3D\cite{bruylants2009jp3d}, which is a 3-D extension of JPEG-2000 and has been widely used as a standard for volumetric image compression.
The traditional wavelet used in JP3D can be implemented by constructing a wavelet basis or decomposing it into basic building blocks, e.g., lifting scheme.
An input image can be converted from the pixel domain to a more energy-concentrated frequency domain through traditional wavelets.
However, there are still many problems in JP3D, which limit its performance. 
Firstly, the traditional wavelet basis in JP3D is designed manually with certain assumptions of the signal, e.g., smooth.
However, volumetric images are usually not ideal and do not follow those assumptions strictly.
Secondly, the traditional wavelet basis performs 1-D filtering in three directions independently, ignoring non-directional correlations. 
Thirdly, the traditional wavelet basis fails to handle various local contexts since it shares wavelet basis functions in different positions of the entire image. 
Last but not least, the traditional wavelet basis employed in JP3D is fixed and cannot be optimized according to characteristics of specific data. 

Video-based methods, such as High Efficiency Video Coding (HEVC), are mainly designed for natural videos. 
They treat one of the three spatial dimensions of volumetric images as a temporal dimension and then estimate a motion field to exploit correlations in that dimension. 
However, in essence, the three spatial dimensions of volumetric images are the same and should be treated equally.
Treating one spatial dimension as a temporal dimension may limit the performance of video-based methods.

Thanks to the vigorous development of deep learning, autoencoder-based methods have developed rapidly and achieved state-of-the-art image compression performance, especially under low bitrate scenarios.
However, autoencoder-based methods employ an irreversible transform module, which leads to uncontrollable losses when converting images into latent features. 
Therefore, they have been shown to have poor performance in the high bitrate scenarios\cite{helminger2020lossy}. 
Unfortunately, high-quality reconstruction is often needed to avoid fatal mistakes in volumetric image compression, and some application scenarios even require lossless reconstruction\cite{bhavani2010survey, kumar2020versatile}. 
To this end, the autoencoder-based compression methods are rarely used in volumetric image compression.  

Motivated by the success of trained 2-D wavelets \cite{ma2019iwave,ma2020end}, we propose a trained 3-D affine wavelet-like transform based on a lifting scheme to overcome the disadvantages of the traditional 3-D wavelet in JP3D.
We replace the prediction and update filters in a lifting scheme with 3-D convolutional neural networks (CNNs).
In this way, our trained 3-D affine wavelet is signal-dependent and can adapt to non-ideal signals. 
In addition, 3-D CNNs make it possible to capture correlations in any direction in 3-D space and can overcome the problems of traditional wavelet basis with only 1-D filters.  
Moreover, compared with using the same wavelet basis globally in traditional wavelet, we introduce the affine wavelet-like transform to adaptively adjust the wavelet basis according to the local content.
We embed the proposed wavelet-like transform into a versatile end-to-end compression framework called aiWave.
We introduce a weight sharing strategy of the affine wavelet-like transform according to the volumetric data characteristics in the axial direction to reduce the encoding and decoding complexities.
Inheriting the advantages of the wavelet transform, our affine wavelet-like transform is reversible and can meet the demand for high reconstruction quality of volumetric images.

Our main contributions are summarized as follows.
\begin{itemize}
	\item We propose a versatile end-to-end compression framework called aiWave, which is composed of trained 3-D affine wavelet-like forward and inverse transforms, entropy coding, and optional quantization and post-processing modules. 
	It can be utilized for both lossy and lossless volumetric image compression.
	\item For the first time, the affine wavelet-like transform is proposed and used for image compression. It is an adaptive and learnable wavelet-like transform that overcomes several shortcomings of the traditional wavelet transform. 
	\item Different weight sharing strategies are explored according to data characteristics to reduce the number of parameters.
\end{itemize}

Part of this work has been published in our previous work \cite{xue2021iwave3d}. 
In this paper, we further propose an affine wavelet-like transform by utilizing an adaptive wavelet basis to overcome the shortcomings of the traditional wavelet.
The weight-sharing strategies are explored under different data characteristics to reduce the complexity of encoding and decoding processes. 
In addition, we propose aiWave-heavy by equipping our affine wavelet-like transform module with a more complex entropy coding module, which finally improves the BD-PSNR by $3.071dB$ compared with our previous work.
Furthermore, the medical image datasets are used to further explore the generalization performance of aiWave in 3-D images.

This paper is organized as follows. 
Section \uppercase\expandafter{\romannumeral2} introduces the related work. 
Section \uppercase\expandafter{\romannumeral3} describes some technical details of our aiWave framework. 
Section \uppercase\expandafter{\romannumeral4} presents our experimental results and analysis. 
Section \uppercase\expandafter{\romannumeral5} concludes the entire paper.

\section{Related Work}
In this section, we give an introduction to the existing methods for volumetric image compression. 
They can be roughly divided into three groups: wavelet-based methods, video-based methods, and autoencoder-based methods. 

\subsection{Wavelet Transform and Its Variants}
\subsubsection{Wavelet transform }
Wavelet transform is highly praised for its excellent decomposition ability. 
It can support both lossy and lossless compression. 
These advantages make it widely used in the field of volumetric image compression\cite{ravichandran2016performance,wang1996medical,selvi2017ct}. 
One representative method based on wavelet transform is JP3D. 
It is the tenth part of JPEG-2000, an extension for 3-D images. 
It is the most widely used method for volumetric image compression. 

By converting the filtering operations into banded matrix multiplications, lifting \cite{sweldens1998lifting,daubechies1998factoring,sweldens1996lifting} scheme is proposed as another way to implement wavelet transform. 
Based on it, Srikal \emph{et al.}\cite{srikala2012neural} proposed a neural network-based compression approach with a lifting scheme.
In \cite{ginesu2004lossy}, the necessity of unitary transform by integer lifting scheme was discussed, and the performance of different integer filter kernels was compared. 
Rossinelli \emph{et al.}\cite{rossinelli2020high} proposed a new lossless data compression scheme based on lifting.

However, all traditional 2-D wavelet transforms above fail to handle 2-D singularities in images \cite{bruylants2015wavelet}. 
To this end, lines and curves in images are ignored when performing traditional wavelet transform.
This problem stems from the fact that the 2-D wavelet transform performs a 1-D discrete wavelet transform in each direction separately from the 2-D image. 
To this end, the traditional wavelet transform cannot capture the non-directional features. 
Moreover, it utilizes these 1-D kernels on the entire image and cannot capture rich local features such as edges and textures. 

\subsubsection{Variants of Wavelet Transform}
In order to solve the problems mentioned above, two kinds of variants of traditional wavelet transform have been proposed.
One kind of method includes ridgelet transform, curvelet transform, and contourlet transform, aiming to solve these problems theoretically. 
The ridgelet transform \cite{do2003finite} was developed to capture straight lines in any direction. 
The key idea was to convert straight lines into several points by radon transform, and then a one-dimension wavelet transform was employed.
Curvelet transform \cite{ma2010curvelet} was actually a block ridgelet transform to handle curves, which was a more general form compared with ridgelet transform. 
Contourlet transform \cite{do2005contourlet} had a more flexible multi-resolution, partial, and reverse image expansion. 
It solved the problems above by performing a directional transform after the subband transform.
Note that curvelet and contourlet transform may produce redundant coefficients and thus are unsuitable for compression tasks.

The other kind of method addresses this problem by adaptively selecting filtering directions. 
Chang \emph{et al.} \cite{chang2007direction} proposed a direction-adaptive wavelet transform based on adaptive directional lifting (ADL). 
Their method predicted the local adjustment of the filtering direction based on image content. 
Liu \emph{et al.} \cite{liu2008weighted} then proposed a new wavelet transform based on weighted adaptive lifting (WAL), which was an improvement of ADL. 
It addressed the mismatch between prediction and updating steps. 

Although these methods improve the traditional wavelet to a certain extent, there are some unresolved problems.
First, traditional wavelet transforms are manually designed for ideal signals instead of images. 
Second, the traditional wavelet basis is fixed when facing different contents.  
These problems affect the performances of traditional wavelet transforms, which prompt us to find a better solution.
Most recently, Ma \emph{et al.} \cite{ma2019iwave,ma2020end} designed a trained wavelet-like reversible transform for end-to-end image compression.
Although it is data-dependent, it still uses a fixed wavelet-like basis when facing different contents.

\subsection{Video-based methods}
Video-based methods regard a certain dimension of volumetric images as the time dimension.
Previous work mostly adjusts the traditional video codec for compressing volumetric images. 
The bit depth ranges of volumetric images are from $8$ to $64$. 
However, due to complexity limitations, video coding codecs often limit the maximum number of coding bits to $16$. 

To solve this problem, Parikh \emph{et al.}\cite{parikh2017high} proposed a high bit-depth medical image compression method using HEVC.
Experimental results show that compared with JPEG-2000, using HEVC can improve compression performance by more than $54\%$.
Some other methods aim to improve the lossless compression of HEVC. 
An improvement to the HEVC intra-frame coding process was proposed in \cite{sanchez2014lossless} for the lossless compression of gray-scale anatomical medical images, which is characterized by a large number of edges. 
Guarda \emph{et al.}\cite{guarda2017method} improved the lossless coding of volumetric medical images for HEVC.
Based on Least Squares Prediction (LSP), a new pixel-by-pixel prediction scheme was explored to extend the current HEVC lossless tools. 

Although these methods can bring noticeable performance improvements, we would still consider that using video-based methods to compress 3-D images is counter-intuitive.
The three dimensions of volumetric images should be treated equally as they are essentially the same.
The time dimension should not be distinguished and treated differently. 

\subsection{Autoencoder-based methods }
Learning-based methods have attracted widespread attention in recent years.
Balle \emph{et al.} \cite{balle2016end} first proposed an end-to-end fully CNN-based compression method. 
Then they introduced hyperprior \cite{balle2018variational} and auto-regressive prior \cite{minnen2018joint} to their methods. 
Among them, \cite{minnen2018joint} was reported to outperform BPG for the first time.
The common part of these popular learning-based methods is that they use an autoencoder as a transform module. 
However, the autoencoder module is irreversible and causes loss of information. 
Some work \cite{helminger2020lossy} reported that these popular methods became convergent at high bitrates, and the performance does not increase as the bitrate increases. 

Gao \emph{et al.} \cite{gao2020volumetric} converted the method mentioned in \cite{minnen2018joint} into its 3-D version to compress brain images. 
This is also the first time that learning-based methods have been used to compress 3-D images. 
Because of the use of a 3-D autoencoder, their method only surpasses JP3D on very low bitrates. 
As we know, high-quality reconstruction is essential for volumetric images. 
Therefore, supporting high-quality reconstruction is an essential requirement for volumetric image compression, which is quite different from natural image compression. 


\section{Proposed Framework}
This section introduces the proposed versatile volumetric image compression framework aiWave in detail. 
First, we give an overview of the structure and pipeline of aiWave. 
Second, we focus on introducing the affine wavelet-like transform module. 
Third, we discuss the weight-sharing strategies of the proposed transform module, followed by the entropy coding and post-processing modules.
Finally, we briefly introduce the loss function to train the whole framework.

\begin{figure}
\centerline{\includegraphics[width=\columnwidth]{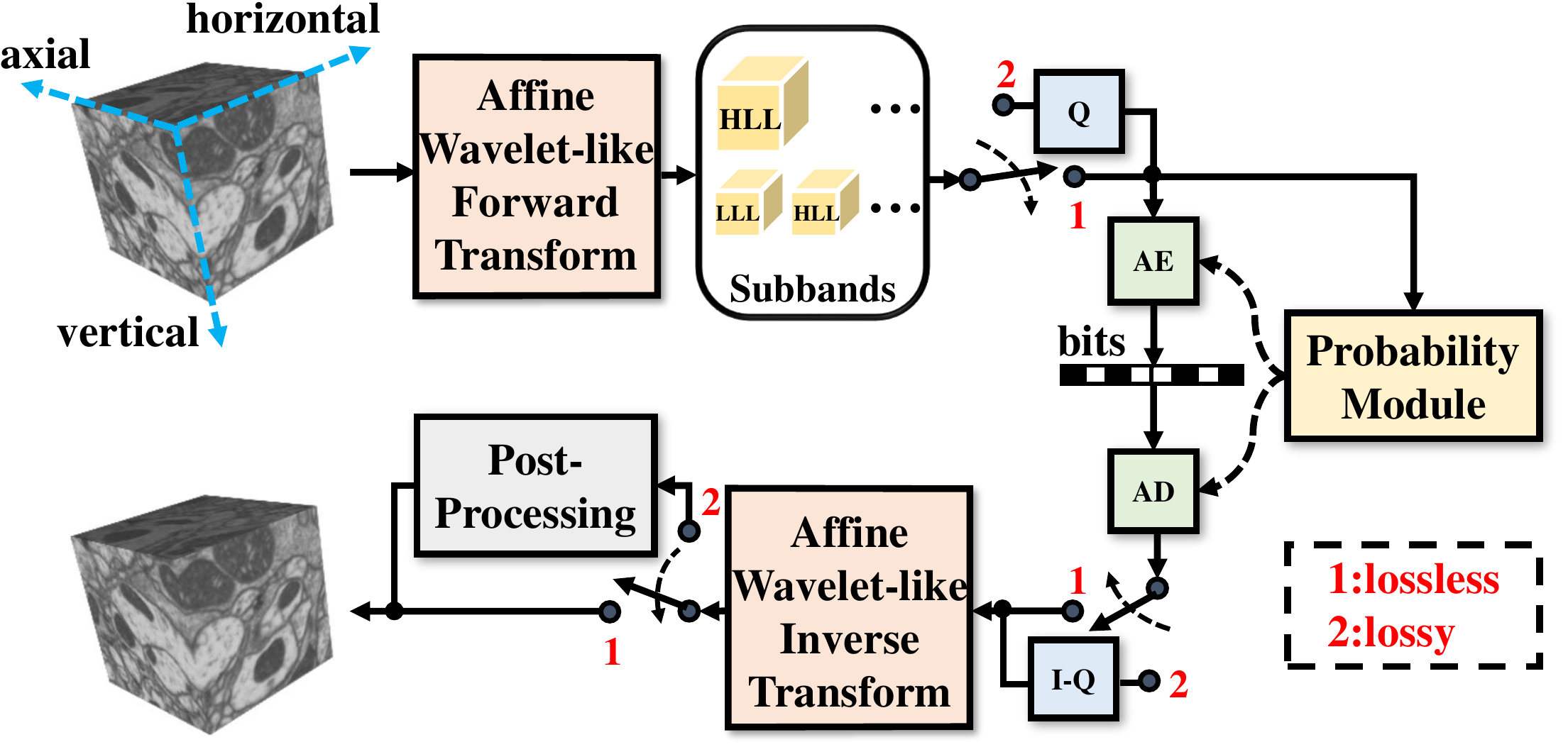}}
\caption{Overview of our proposed aiWave framework. ``Q" and ``I-Q'' stand for quantization and inverse-quantization module, ``AE" and ``AD" stand for arithmetic encoding and arithmetic decoding module, respectively. The affine wavelet-like forward transform and inverse transform modules are implemented with 3-D affine lifting operations (see Fig. \ref{fig-lifting} for the description of 1-D affine lifting operations). The details of the two kinds of entropy coding modules we explored are shown in Fig. \ref{Entropy model}. The post-processing module is utilized to compensate for quantization errors. aiWave can support lossless compression by turning the switch to 1 and lossy compression by turning the switch to 2.}
\label{fig1}
\label{Overview}
\end{figure}

\subsection{Overview of the framework}
Fig.~\ref{Overview} shows our proposed versatile volumetric image compression framework aiWave. 
It mainly includes four main modules: affine wavelet-like forward transform module, entropy coding module, affine wavelet-like inverse transform module, and post-processing module. 
The overall encoding and decoding processes are shown as follows. 
 
In the encoding process, the input image \(x \in {\mathbb{R}^N}\) is converted into a series of subband coefficients \(y\):
\begin{equation}
y = {g_a}(x;{\phi}),
\end{equation}
where $g_a(x;\phi)$ is affine wavelet-like forward transform implemented with 3-D CNNs.
Then $y$ is quantized and rounded to its nearest integer: 
 \begin{equation}
q =\left[ {y/QS} \right],
\end{equation}
where [$\cdot$] is the rounding operation and $QS$ is the quantization step.
It is worth noting that the transform is reversible and does not lead to any loss of information.
Only the quantization process introduces loss in the entire framework. 
Next, the quantized coefficients \(q\) are sent to the arithmetic encoder to generate the bitstream. 
An entropy coding module is used to estimate the probability distribution of coefficients to assist the arithmetic encoder. 
This paper explores two entropy coding modules: a light one implemented by factorized entropy model and a heavy one to further remove redundancy.

In the decoding process, the bitstream is first decoded by an arithmetic decoder with the help of the entropy model.
Then, the decoded coefficients are inversely quantized to obtain the coefficient \(\hat y\):
\begin{equation}
\hat y = q \times QS.
\end{equation}
Finally, the restored image $\hat{x}$ is obtained by
\begin{equation}
\hat x = {g_a^{-1}}(\hat y;\phi ).
\end{equation}
Note that affine wavelet-like inverse transform $g_a^{-1}$ is the inverse of affine wavelet-like forward transform $g_a$.
Finally, \(\hat x\) is enhanced by the post-processing module to remove compression noise.

The quantization, inverse-quantization, and post-processing modules are removed when performing lossless compression.
At the same time, the transform module should also be changed to its lossless version. 

\subsection{Affine wavelet-like transform module}
\label{affine section}
\begin{figure}\textbf{}
\centering
\subfigure[ Traditional 1-D lifting scheme]
{
\includegraphics[width=0.47\textwidth]{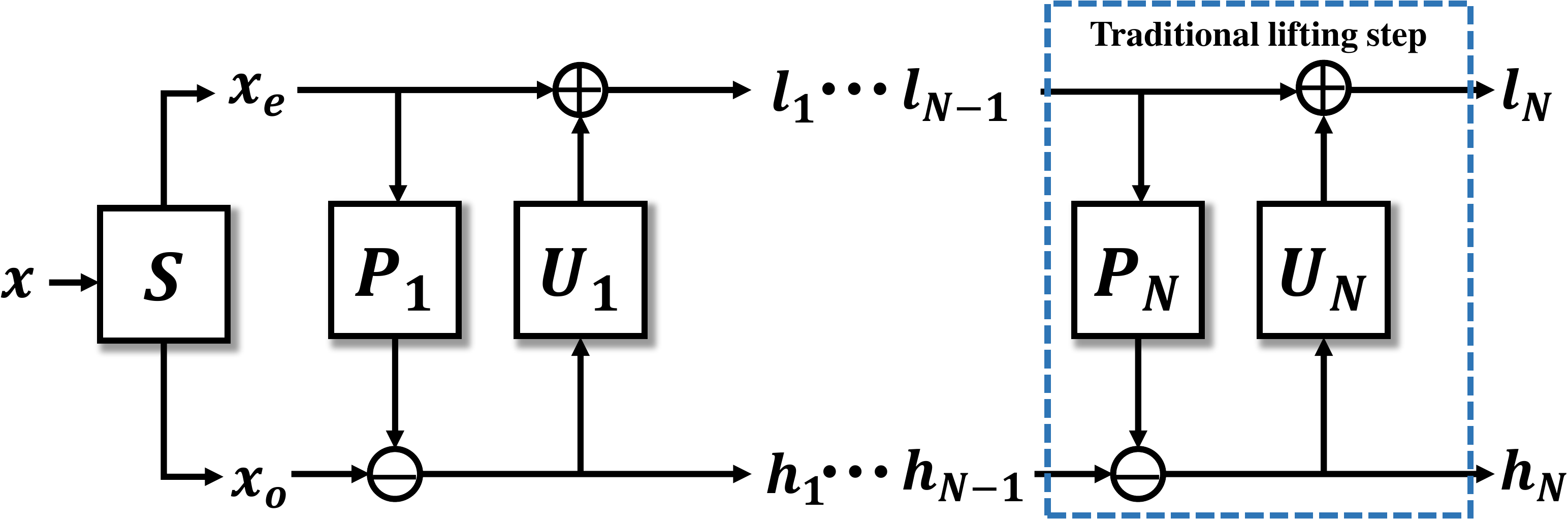}
}
\subfigure[Proposed 1-D affine wavelet-like transform]
{
\includegraphics[width=0.47\textwidth]{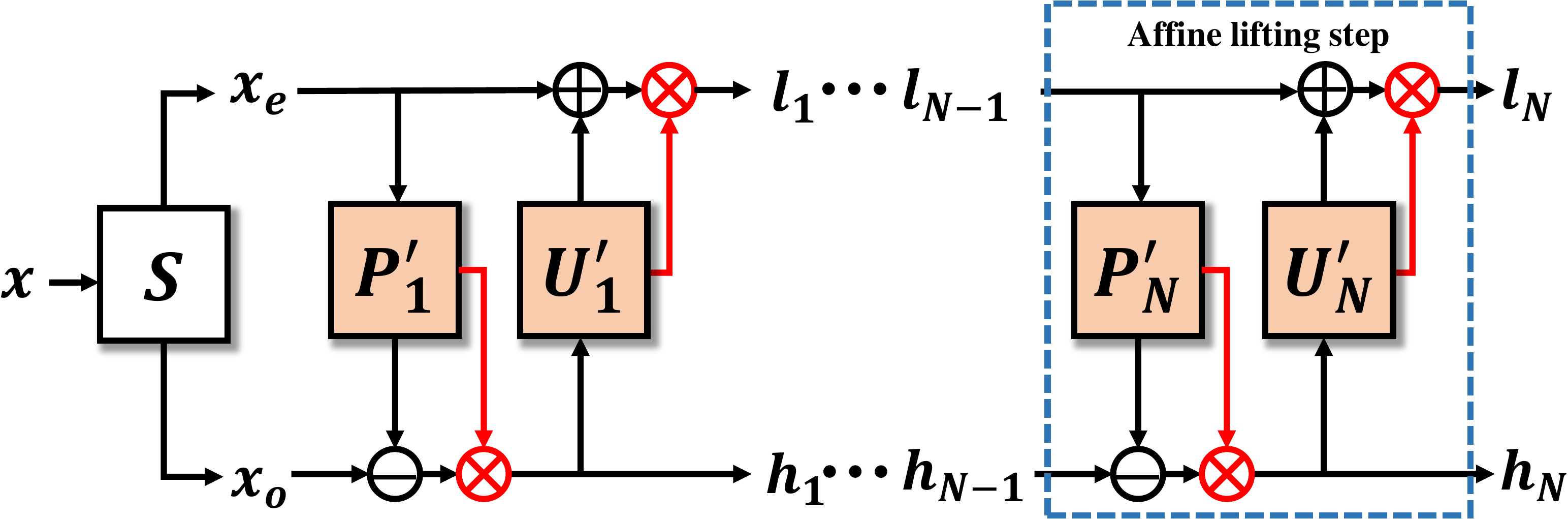}
}
\caption{Difference between the traditional 1-D lifting scheme and the proposed 1-D affine wavelet-like transform. \(P_{_i}\) stands for prediction filter, and \(U_{_i}\) stands for update filter. 
\({P_i}\)' stands for prediction network and \({U_i}\)' stands for update network. The main difference between the two is that the affine wavelet-like transform additionally predicts an affine map and uses 3D-CNNs to replace the traditional filters. N is the total number of lifting steps in the 1-D transform. We use N=2 in this paper. For volumetric images, the transform is performed three times, first axial-wise, then horizontal-wise, and finally vertical-wise.}
\label{fig-lifting}
\end{figure}
\subsubsection{Structure of affine wavelet-like transform}

Our proposed affine wavelet-like transform is built based on a traditional lifting scheme. 
Fig.~\ref{fig-lifting} shows the difference between the 1-D traditional lifting scheme and the 1-D affine wavelet-like transform.

The traditional lifting scheme includes three operations, namely split, prediction, and update. 
Through these operations, the lifting scheme aims to decompose the original signal into high-frequency and low-frequency components.
One prediction operation and one update operation constitute a basic lifting step. 
A split operation incorporating N basic lifting steps forms a 1-D traditional lifting scheme.

We take a 1-D signal as an example to show the detailed lifting scheme.
First, the input signal is split into an odd component \(x_{_o}\) and an even component \(x_{_e}\),
\begin{equation}
{x_o},{x_e} = split(x).
\end{equation}
Note that there is a strong similarity between \(x_{_o}\) and \(x_{_e}\).
Therefore, the prediction operation is performed to predict \(x_{_o}\) from \(x_{_e}\),
\begin{equation}
h = {x_{_o}} - {P_i}({x_e}),
\label{equation 6}
\end{equation}
where \(P_i\) is the prediction filter, and \( h\) represents the prediction residual containing high-frequency information of \(x\). 
Then, \(h\) is utilized to update \(x_{_e}\) through the update operation,
\begin{equation}
l = {x_e} + {U_i}(h),
\end{equation}
where \({U_i}\) is the update filter, and \( l\) contains the coarse details in \( x\), that is, low-frequency information.  
Note that \({P_i}\) and \({U_i}\) in traditional wavelet transform are simple linear filters. 
For example, in the \(CDF\) 5/3 wavelet, which is used in JP3D for lossless compression, \eqref{equation 6} can be written as: \(h[m] = {{\rm{x}}_o}[m] - ({p_a}{{\rm{x}}_e}[m - 1] + {p_b}{{\rm{x}}_e}[m])\), where \({p_a} = {p_b} = 0.5\). 
Different prediction and update filters correspond to different wavelet transforms\cite{daubechies1998factoring}. 

\begin{figure}
\centerline{\includegraphics[width=0.84\columnwidth]{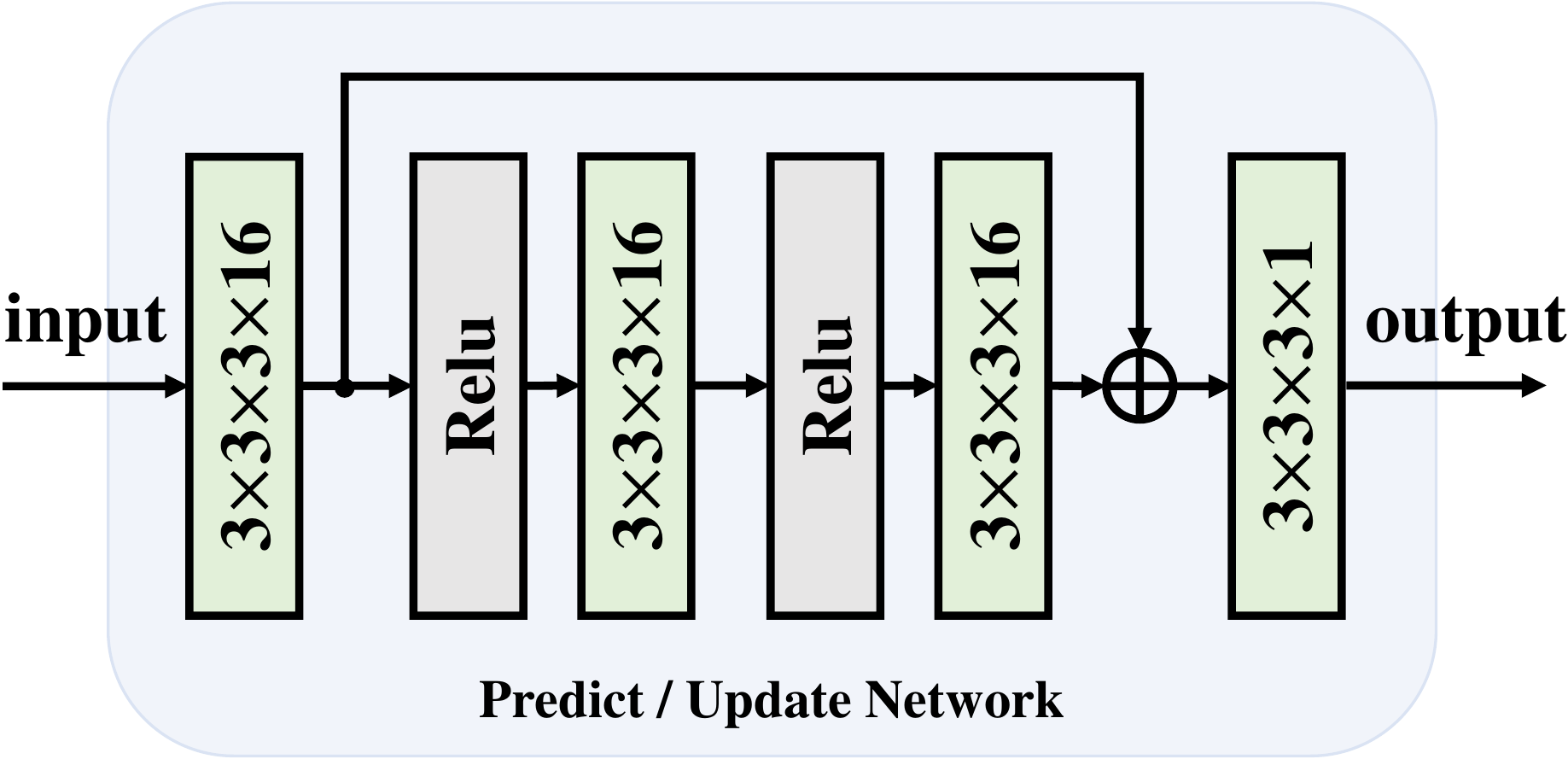}}
\caption{3-D convolutional neural network used for each prediction or update step (see Fig. \ref{fig-lifting}). The shown numbers like $3\times3\times3\times16$ indicate the kernel size ($3\times3\times3$) and the number of output channels (\(16\)) of each layer. ``Relu'' indicates the adopted nonlinear activation function.}
\label{Network structure}
\end{figure}

Considering that the above prediction and update filters are designed manually, in this paper, we try to replace these filters with 3-D CNNs.
The structure of our proposed 3-D CNNs is shown in Fig.~\ref{Network structure}. 
In addition, inspired by the generative model glow\cite{dinh2014nice,dinh2016density,kingma2018glow}, we further predict an affine map to adjust the wavelet basis according to the context as shown in Fig.~\ref{fig-lifting} (b).
Therefore, the proposed affine lifting step can be written as:
\begin{equation}
{x_o},{x_e} = split(x),
\end{equation}
\begin{equation}
h = {A({x_e})\odot (x_{_o}} - {P_i}'({x_e})),
\label{equ-9}
\end{equation}
\begin{equation}
l = {B(h) \odot ({x_e}} + {U_i}'(h)),
\label{equ-10}
\end{equation}
where \({P_i}'\)and \({U_i}'\) represent the prediction and update network, respectively. 
\(A\) and \(B\) represent the network for predicting the affine map, which has the same structure as \({P_i}'\)and \({U_i}'\), except that a sigmoid operation is added to limit the values in the affine map to the range between 0 and 1.
Similar to the traditional lifting scheme, \eqref{equ-9} and \eqref{equ-10} form a basic affine lifting step. 
A split operation incorporating N basic affine lifting steps forms a 1-D affine wavelet-like transform. 

\begin{figure}
\includegraphics[width=\columnwidth]{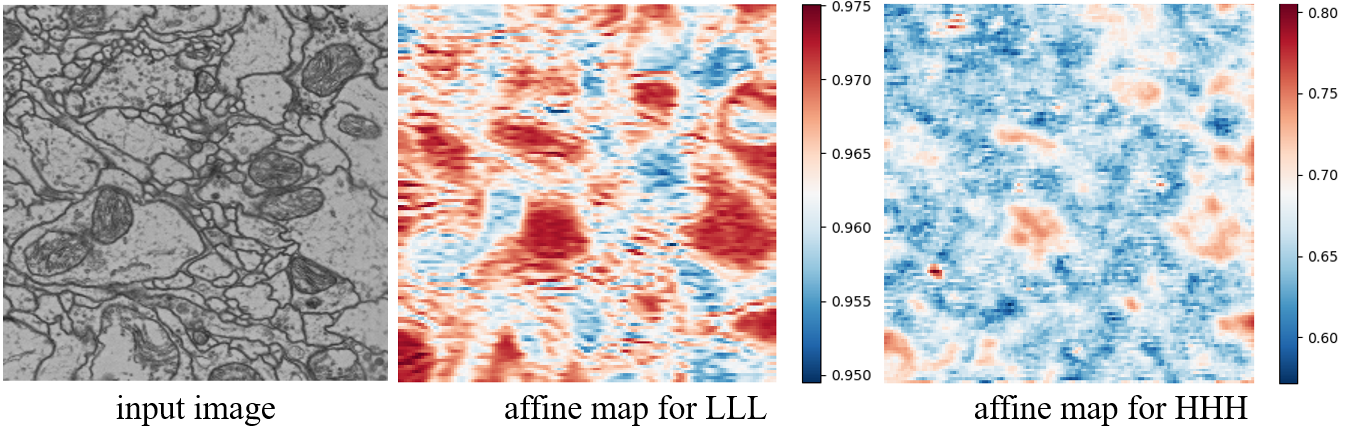}
    \caption{Visualization results of affine maps. Affine maps for the LLL subband and HHH subband are given. It can be seen that the value of the affine map will change as the content of the input image changes. Moreover, the affine value range of different subbands is very different. We notice that the value of the affine map of the subband LLL is much larger than the value of the subband HHH on average.}
    \label{affine_visual}
\end{figure}

In the affine wavelet-like transform, the affine map predicts a value for each spatial position. 
We call it a fine-grained affine wavelet-like transform. 
As shown in Fig.~\ref{affine_visual}, after network training, we find out that the difference between the affine maps is much more significant than that between the different positions of the same affine map. 
This motivates us to simplify each affine map in the affine wavelet-like transform into a trainable variable, and we call it a coarse-grained affine wavelet-like transform. 
Results of fine-grained and coarse-grained affine wavelet-like transform are further discussed in Section~\ref{fine-coarse}.

When transforming a 3-D image using the proposed 3-D wavelet-like transform, we first perform a 1-D affine wavelet-like transform in the axial direction of the 3-D image. 
After that, the original image is decomposed into a low-frequency subband \textit{\(L\)} and a high-frequency subband \textit{\(H\)}. 
Then we perform a 1-D affine wavelet-like transform on \(L\) and \(H\) in the horizontal direction, resulting in four subbands \{\( {LL, HL, LH, HH}\)\}. 
For the resulting four subbands, finally, another 1-D affine wavelet-like transform is carried out in the vertical direction, and eight subbands \{$LLL$, $HLL$, $LHL$, $HHL$, $LLH$, $HLH$, $LHH$, $HHH$\} are obtained. 
Three such 1-D affine wavelet-like transforms constitute a complete decomposition. 
The LLL subband can be further decomposed into eight subbands to make the energy more concentrated. 
After N times of decomposition, 7N+1 subbands are obtained. 
Note that the parameters of affine wavelet-like transforms in different decomposition levels are shared in our method. 
To this end, the scale-invariant characteristics of traditional wavelets are still satisfied in our affine wavelet-like transform.

In addition, our affine wavelet-like transform is reversible. 
The affine wavelet-like inverse transform can be easily obtained from the forward transform as follows: 

\begin{equation}
{x_e} = l / B(h) - {U_i}'(h),
\end{equation}
\begin{equation}
{x_o} = h/A({x_e}) + {P_i}'({x_e}),
\end{equation}
\begin{equation}
x = M({x_e},{x_o}),
\end{equation}
where \(M\) means the merge operation, which is a reverse operation of split.
Note that \({P_i}'\), \({U_i}'\), \(A\) and \(B\) in inverse transform share parameters with that in forward transform.

\subsubsection{Lossless affine wavelet-like transform}
To guarantee lossless compression in a traditional lifting step, a rounding operation is usually added to the prediction and update operations,
\begin{equation}
h = {x_{_o}} - \left [{P_i}'({x_e}) \right],
\label{equ-14}
\end{equation}
\begin{equation}
l = {x_e} + \left[ {U_i}'(h) \right].
\label{equ-15}
\end{equation}
In this way, the entire traditional lifting scheme has only integer operations, thus avoiding loss of precision.
However, the situation is different in the proposed affine lifting step. 
Our method introduces multiplication and division operations, which can bring precision losses, especially in hardware implementation. In this paper, for simplification, we propose a lossless compression scheme by setting each position of the affine map to a constant 1. 
At this time, the lossless scheme of aiWave is consistent with \eqref{equ-14} and \eqref{equ-15}.

\begin{figure*}
\centerline{\includegraphics[width=2\columnwidth]{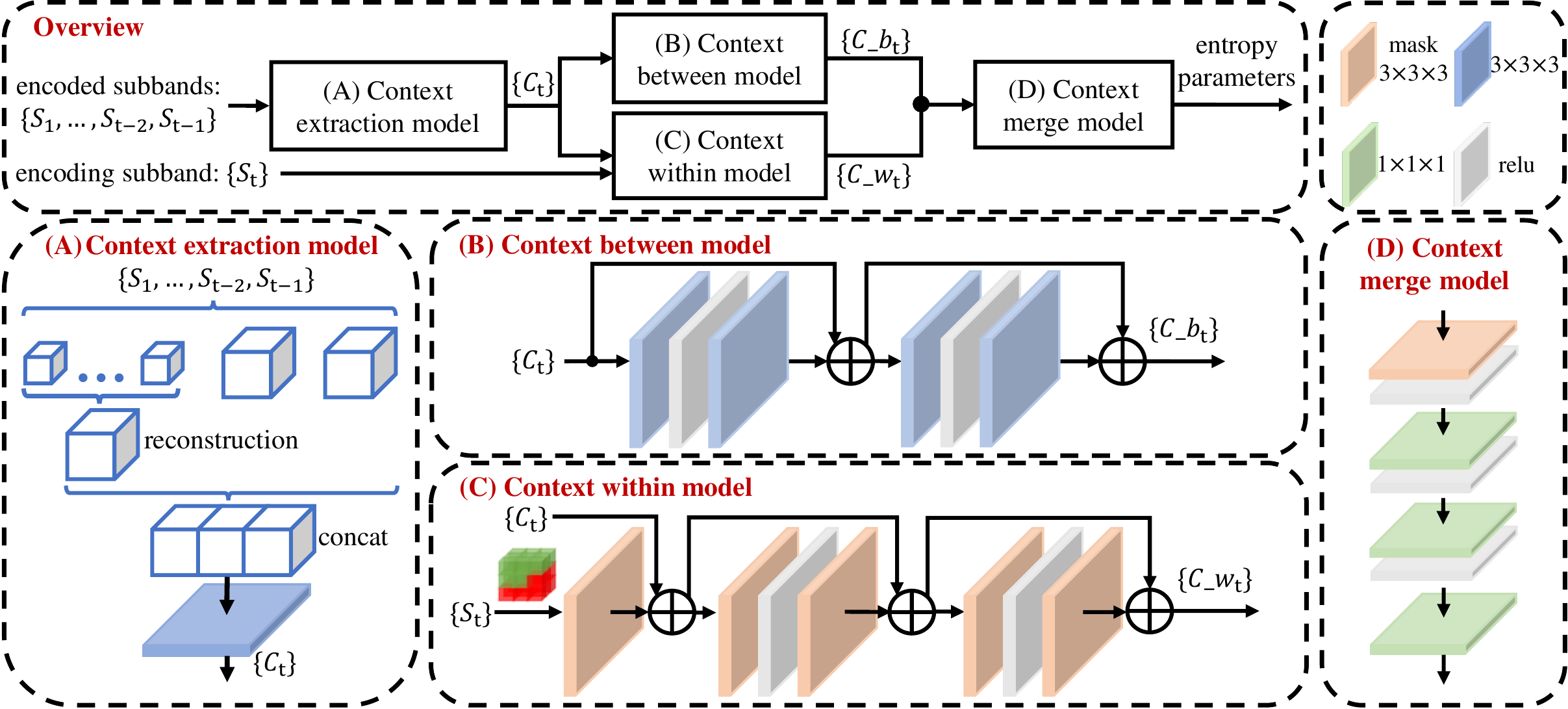}}
\caption{Structure of context-based entropy coding module. $3\times3\times3$ represents the kernel size of CNN, and mask $3\times3\times3$ represents the kernel size of Mask CNN. The main idea of this module is to use the correlation between and within subbands to reduce the uncertainty of the currently coded pixels, thereby reducing the bitrate. Firstly, the context extraction model roughly extracts the correlation between subbands(\(C_t\)), and \(C_t\) is further input into context between model and context within model to extract the context between and within subbands(\(C\_{b_t}\) and \(C\_{w_t}\)). Then context merge model is utilized to fuse these context information and obtains entropy parameters.}
\label{Entropy model}
\end{figure*}

\subsubsection{Theoretical analysis of affine wavelet-like transform}
In this section, we theoretically explain that the proposed affine wavelet-like transform improves the wavelet basis of the traditional wavelet transform into a data-driven and content-adaptive form, thus improving the performance. 

According to \cite{daubechies1998factoring}, the traditional wavelet transform for finite impulse response filter can be decomposed into a lifting scheme by performing factorization. 
For example, the z-transform of a analysis filter pair \((h, g)\) is denoted by \(h(z)\) and \(g(z)\), which represent the low-pass and high-pass filter, respectively. Then the corresponding decomposition polyphase matrix \(P(z)\) is defined as follows:
\begin{equation}
P(z) = \left[ {\begin{array}{*{20}{c}}
{{h_e}(z)}&{{h_o}(z)}\\
{{g_e}(z)}&{{g_o}(z)}
\end{array}} \right]
\end{equation}
where the \({h_e}(z)\) and \({g_e}(z)\) are the even parts and the \({h_o}(z)\) and \({g_o}(z)\) are the odd parts of the low-pass and high-pass filter.
The polyphase matrix \(P(z)\) can be factorized into: 
\begin{equation}
P(z) = \prod\limits_{i = 1}^m {\left[ {\begin{array}{*{20}{c}}
1&{{s_i}(z)}\\
0&1
\end{array}} \right]\left[ {\begin{array}{*{20}{c}}
1&0\\
{{t_i}(z)}&1
\end{array}} \right]\left[ {\begin{array}{*{20}{c}}
K&0\\
0&{1/K}
\end{array}} \right]}.
\end{equation}
Taking the CDF 9/7 filter bank used in JP3D as an example, the CDF 9/7 filter pair can be written as: 
\begin{equation}
\resizebox{1.0\hsize}{!}{$P(z)\! = \left[ {\begin{array}{*{20}{c}}
1\!&{\alpha f(z)\!}\\
0\!&1\!
\end{array}} \right]\left[ {\begin{array}{*{20}{c}}
1\!&0\!\\
{\beta g(z)\!}&1\!
\end{array}} \right]\left[ {\begin{array}{*{20}{c}}
1\!&{\gamma f(z)\!}\\
0\!&1\!
\end{array}} \right]\left[ {\begin{array}{*{20}{c}}
1\!&0\!\\
{\delta g(z)\!}&1\!
\end{array}} \right]\left[ {\begin{array}{*{20}{c}}
\zeta\! &0\!\\
0\!&{1\!/\zeta\! }
\end{array}} \right]$},
\end{equation}
where \(f(z) = 1+{z^{-1}}\), \(g(z) = 1+z\),  \(\alpha  \approx  - 1.586\), \(\beta  \approx  - 0.053\), \(\gamma  \approx 0.883\), \(\delta  \approx 0.444\),  \(\zeta  \approx  - 1.150\).
As we know, \(\alpha\), \(\beta\), \(\gamma \) and \(\delta\) are the parameters of prediction and update filters in a traditional lifting scheme.
\(\zeta\) is the scaling factor, which can achieve overall scaling for the previous parameters. 
Therefore, changing these parameters in the lifting scheme corresponds to the selection of different wavelet basis in the first generation of wavelets. 

In the affine wavelet-like transform, we first replace the prediction and update filters with 3D CNNs. 
This indicates that we improve the traditionally handcrafted wavelet basis to a wavelet-like basis learned from data. 
More importantly, once the traditional wavelet basis is determined, it does not change when used. 
However, our affine wavelet-like transform also predicts an affine map to form the wavelet-like basis, which varies according to input content.
In other words, our affine wavelet-like transform adaptively learns a pixel-wise wavelet basis for each different spatial position, similar to a spatial attention mechanism. 

\subsection{Weight sharing strategy }
\label{weight}

It is well known that the physical properties of 2-D natural images in horizontal and vertical directions are the same.
Therefore, the traditional wavelet, designed for 2-D images, share parameters in horizontal and vertical directions. 
However, most 3-D images do not satisfy this property. 
This phenomenon is caused by the technical limits of the image scanner.
Although imaging technologies try to achieve the same resolution in the axial direction (\(z\)) as in the lateral direction (\(x\), \(y\)), most of them, such as serial section transmission EM (ssTEM) fail to collect such thin slices. 
Thus the image resolution within a slice is much higher than between slices in most cases, which means that the correlation in the axial dimension is much smaller than that in the other two dimensions. 
A few EM, such as focused ion beam scanning EM (FIB-SEM), can collect isotropic data with the same resolution in all three directions through layer-by-layer corrosion samples.

This inspires us to explore whether the parameter sharing strategy should be adjusted according to the characteristics of the 3-D images. 
For isotropic data, we propose to share parameters of the prediction and update the network in all directions. 
For anisotropic data, we use a single set of parameters of 3D CNNs to learn wavelets adapted to their characteristics in the axial direction and share parameters in the other two directions. 
The detailed experimental results are discussed in Section \ref{weight results}. 

\begin{table*}
\centering
\caption{Datasets Introduction And Training Details}
\scalebox{1.06}
{
\begin{tabular}{lccccccc}
\toprule[0.8pt]
Name &Category&Species&Position &Characteristic &Bit depth&w/o sign &Resolution\\
\toprule[0.8pt]
FAFB\cite{zheng2018complete}  &biological  &adult drosophila &brain   &anisotropy &8 &unsigned & -\\ 
FIB-25\cite{takemura2015synaptic} & biological &adult drosophila  &brain & isotropy &8&unsigned & 520\(\times\)520\(\times\)520 \\ 
CT-spleen\cite{simpson2015chemotherapy} &medical &human  &spleen  &anisotropy &32&signed & 33$\sim$168\(\times\)512\(\times\)512 \\ 
MRI-heart\cite{tobon2015benchmark} &medical &human  & heart & anisotropy &32&signed & 90$\sim$180\(\times\)320\(\times\)320  \\ 
Chaos-CT\cite{kavur2021chaos} &medical &human  &abdominal  &anisotropy &16&signed & 78$\sim$294\(\times\)512\(\times\)512\\ 
Attention\cite{buchel1997modulation} &medical &human  & brain & anisotropy &16&unsigned & 44\(\times\)60\(\times\)52 \\ 
\bottomrule[0.8pt]
\end{tabular}
}
\label{datasets details}
\end{table*}

\subsection{Entropy coding module}
In this paper, to adapt different application scenarios, two kinds of entropy coding modules are employed to estimate the probability distributions of the quantized subband coefficients.

\subsubsection{Factorized entropy coding module}
For scenarios requiring low latency, the factorized entropy coding module is utilized in our framework, and we call this framework aiWave-light.  

After transform and quantization, a series of quantized subband coefficients are obtained. 
The factorized entropy model\cite{balle2018variational} estimates a continuous probability distribution for each subband. 
These probabilities can then be used to assist the arithmetic coding process. 
It is worth mentioning that the subbands are considered independent and are coded and decoded in parallel in aiWave-light. 
Therefore, aiWave-light has low coding and decoding complexities and is suitable for low latency scenarios. 

\subsubsection{Context-based entropy coding module}
For scenes requiring high performance, we employ a context-based entropy coding module to further remove redundancies within and between subbands. 

As shown in Fig. \ref{Entropy model}, our context-based entropy coding module mainly includes four sub-modules. 
First, the context extraction model is used to roughly obtain the context of the current subband.
Then, we employ the context between model and context within the model to further extract context information between different subbands and within the same subband through the coded pixels.
Finally, a context merge model is employed to fuse these context information and get a series of entropy parameters for probability estimation. 

\textbf{Context extraction model: }We employ the context extraction model to roughly extract the context \(C_t\) from the coded subbands \{\(S_1, S_2,..., S_{t-1}\)\}. 
\begin{equation}
{C_{\text{t}}} = Conv(Recon({S_1},{S_2},...,{S_{t - 1}})).
\end{equation}
\(C_t\) is further used in context between model and context within model. 
Note that if the resolution of \(S_{t - 1}\) is not the same as that of \(S_{t}\), we reconstruct a high-resolution low-frequency subband \( LLL \) with the affine wavelet-like inverse transform. 
After all the coded subbands are concatenated, a 3-D convolution is applied to obtain \(C_t\).

\textbf{Context between model: } We employ the context between model to remove redundancies between the current subband and the coded subbands.
\begin{equation}
C\_{b_t} = B({C_t}).
\end{equation}\(B( \cdot )\) means two cascaded residual blocks used to extract high-dimensional features \(C\_{b_t}\) from \(C_t\).

\textbf{Context within model:} Motivated by PixelCNN\cite{oord2016conditional}, we utilize a context within model to remove redundant information in the same subband. 
\begin{equation}
C\_{w_t} = C({C_t},{S_t}).
\end{equation}\(C( \cdot )\) is the masked 3-D convolutional layers used to extract high-dimensional features \(C\_{w_t}\) from \(C_t\) and \(S_t\).

\textbf{Context merge model: } After extracting the context \(C\_{b_t}\) and \(C\_{w_t}\), they are concatenated  together in the channel dimension. Then the concatenated result is input to the context merge model to calculate a series of entropy parameters \(\psi\).

\begin{equation}
\psi = D(concat(C\_{b_t},C\_{w_t})).
\end{equation}
Note that the output \(\psi\) has 58 channels and different channels of \(\psi\) corresponding to different entropy parameters in a cumulative probability distribution, which is defined as follows:

\begin{equation}
c = {f_K} \circ {f_{K - 1}} \cdot  \cdot  \cdot {f_1}.
\end{equation}
Symbol \(\circ\) means the composition of functions. That is to say, the cumulative probability distribution function is a composition of functions \(f_k\). K is set to 5 in our experiments. For \(k=1,\) \(f_k\) is: 

\begin{equation}
{f_1}(\textbf{x}) = {g_1}({\textbf{H}^1}\textbf{x} + {\textbf{b}^1}),
\end{equation}
\begin{equation}
{g_1}(\textbf{x}) = \textbf{x} + {\textbf{a}^1} \cdot \tanh (\textbf{x}),
\end{equation}
where \({\textbf{H}^1},{\textbf{b}^1},{\textbf{a}^1} \in {\mathbb{R}^{3 \times 1}}\). When \(1< k < K\),

\begin{equation}
{f_k}(\textbf{x}) = {g_k}({\textbf{H}^k}\textbf{x} + {\textbf{b}^k}),
\end{equation}
\begin{equation}
{g_k}(\textbf{x}) = \textbf{x} + {\textbf{a}^k} \cdot \tanh (\textbf{x}),
\end{equation}
where \(\textbf{H}^k\in{\mathbb{R}^{3 \times 3}}\) , \(\textbf{b}^k\), \(\textbf{a}^k \in{\mathbb{R}^{3 \times 1}}\). The dot operation is element-wise multiplication. When \(k=K\),
\begin{equation}
{f_K}(\textbf{x}) = sigmoid({\textbf{H}^K}\textbf{x} + {b^K}),
\end{equation}
where \(\textbf{H}^K \in{\mathbb{R}^{1 \times 3}}\), \({b}^K \in{\mathbb{R}}\).

All of the \(\textbf{H}^k \in{\mathbb{R}^{1 \times 3}}\), \(\textbf{b}^k/b^K\), and \(\textbf{a}^k\) above constitute the entropy parameters \(\psi\). 
We get the estimated cumulative probability distribution for the subband coefficients by learning these parameters. 
This cumulative probability distribution is then differentiated to obtain the required probability distribution according to \cite{balle2018variational},
\begin{equation}
p = {f_K^{'}} \cdot {f_{K-1}^{'}} \cdots {f_1^{'}}.
\end{equation}

\subsection{Post-processing module}
As mentioned above, our transform module is reversible.
Therefore, the inverse transform cannot compensate the loss caused by quantization. 
To compensate for quantization loss in aiWave, we introduce a post-processing module after affine wavelet-like inverse transform.
Specifically, we employ multiple 3-D convolutional layers to construct a 3-D residual block, and the post-processing module contains six 3-D residual blocks. 
Note that this is different from autoencoder-based frameworks whose decoders can achieve inverse transform and reconstructed image enhancement simultaneously. 

\subsection{Loss function}
During training, we treat the framework as a rate-distortion optimization problem. Thus the loss function becomes:
\begin{equation}
\label{eq::optimization}
L = R + \lambda  \cdot D = \underbrace {\mathbb{E}{_{x \sim {p_x}}}\left[ { - {{\log }_2}{p_{q}}\left( {q} \right)} \right]}_{{\rm{rate}}} + \lambda  \cdot \underbrace {\mathbb{E}{_{x \sim {p_x}}}\left\| {x - \hat x} \right\|_2^2}_{{\rm{distortion}}},
\end{equation}
where \(R\) and \(D\) represents bitrate and distortion, respectively. q stands for quantized coefficients. 
For each model, we specify a Lagrange multiplier \(\lambda\), which controls the tradeoff between rate and distortion. 
MSE is utilized to compute the distortion between input image \(x\) and reconstruct image \(\hat x\).

\section{Experimental results}
\label{experimental results}

\subsection{Datasets}
In our experiments, six volumetric image datasets are used for extensive quantitative assessment.
Note that under the proposed framework, a different group of network parameters are trained for different datasets.
Detailed information about these datasets  are shown in Table~\ref{datasets details}. 

\textbf{FAFB dataset\cite{zheng2018complete}:}
The FAFB dataset is a adult fly brain electron microscope anisotropic imaging, imaged at 4\(\times\)4\(\times\)40 \(nm^3\) resolution. The complete dataset contains 106TB of volumetric images. In our experiments, we downloaded four 64×768×768 blocks located in the cortical area and cropped them into 64\(\times\)64\(\times\)64 blocks, of which 90\% of the blocks were used for training and the remaining 10\% for testing.

\textbf{FIB-25 dataset\cite{takemura2015synaptic}:}
The FIB-25 dataset is the raw EM data for the 7 medulla column FIB-SEM reconstruction. The complete set includes 8000 images, combined into one large tar file(about 100 GB). FIB-25 is an isotropic dataset that is imaged at 8\(\times\)8\(\times\)8 \(nm^3\) resolution. We exploit the part with segmentation annotations, which have already been divided into training and testing sets. We cropped them into 64\(\times\)64\(\times\)64 blocks, and 80\% of them were used for training, others for testing.

\textbf{CT-spleen dataset\cite{simpson2015chemotherapy}:}
The CT-spleen dataset was used to investigate the relationship between changes in spleen volume after preoperative chemotherapy and postoperative complications. This dataset includes 61 3D images of sizes from 33\(\times\)512\(\times\)512 to 168\(\times\)512\(\times\)512 in nii.gz format. We chose the images with the size of the z-direction more than or equal to 64 and cropped them into 64\(\times\)64\(\times\)64 blocks. Among them, blocks with more than half of the pixels being 0 are discarded. We got 386 64\(\times\)64\(\times\)64 blocks. Sixty-four of them were randomly selected as the testing and the rest for training. 

\textbf{MRI-heart dataset\cite{tobon2015benchmark}:}
The MRI-heart dataset includes 31 3D heart MR images with nii.gz format, the size is from 90\(\times\)320\(\times\)320 to 180\(\times\)320\(\times\)320. After performing the same cropping and dropping as the CT-spleen dataset, we got 283 training blocks and 64 testing blocks with sizes 64\(\times\)64\(\times\)64.
Note that the bit depth of the CT-spleen and MRI-heart dataset is 32.
As we know, JP3D and HEVC do not support the volumetric image compression with input bit depth larger than 16.
To compare aiWave with JP3D and HEVC, we scale the CT-spleen and MRI-heart images to 16 bits by performing min-max normalization following \cite{wang20193d}.

\textbf{Chaos-CT dataset\cite{kavur2021chaos}:}
This dataset contains many DICOM images of abdominal CT and MRI from multiple patients. We select CT images for quantitative assessment. These CT images were obtained in the upper abdominal region following the injection of contrast in the patient's portal venous phase. The size is from 78\(\times\)512\(\times\)512 to 294\(\times\)512\(\times\)512. Again, we cropped it into 64\(\times\)64\(\times\)64 and dropped the blocks with more than half of the pixels being 0. Then we got 978 blocks, and 64 of them were randomly selected for testing.

\textbf{Attention dataset\cite{buchel1997modulation}:}
The Attention dataset was used to resolve attentional modulation of active connectivity using functional magnetic resonance imaging (fMRI).
The original dataset contains 360 images in the HDR format without partition. The size of each image is 44\(\times\)60\(\times\)52. We used the first 300 of them for training and the last 60 for testing.

\begin{figure*}
    \centering
	\subfigure[FAFB dataset]
    {
	\includegraphics[width=0.62\columnwidth]{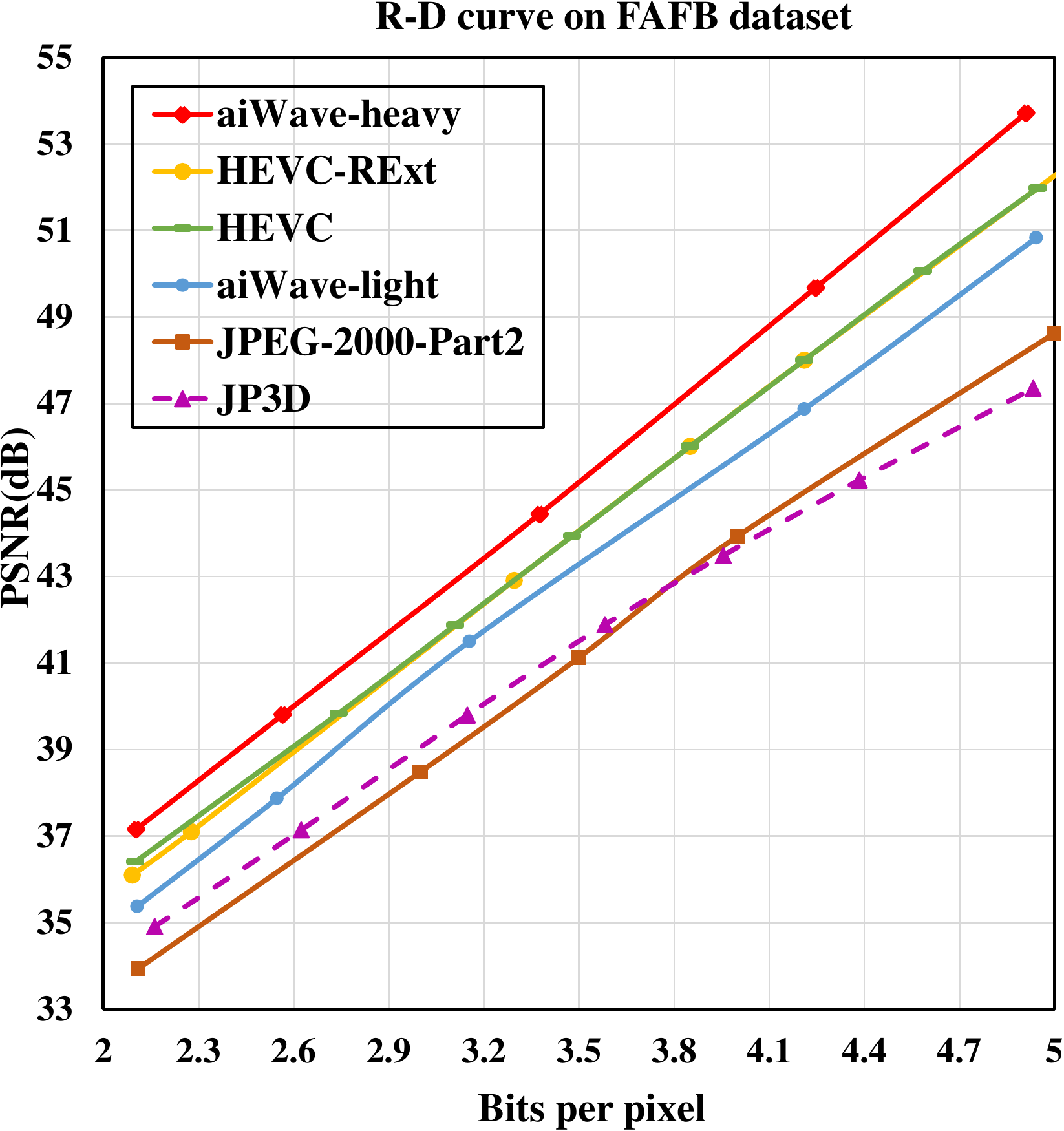}
    }
   \subfigure[FIB-25 dataset]
    {
	\includegraphics[width=0.62\columnwidth]{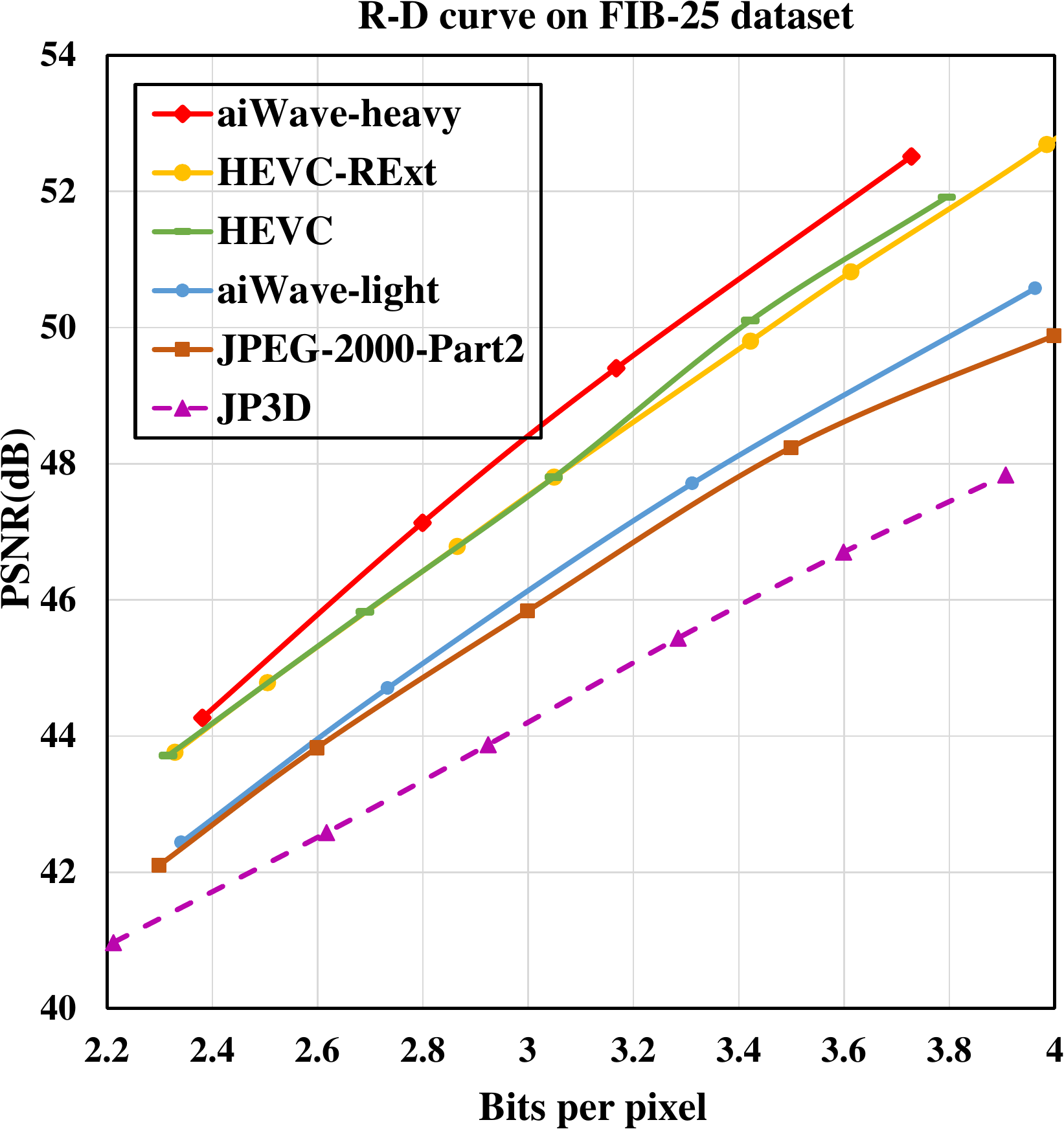}
     }   
    \subfigure[CT-Spleen dataset]
    {
	\includegraphics[width=0.63\columnwidth]{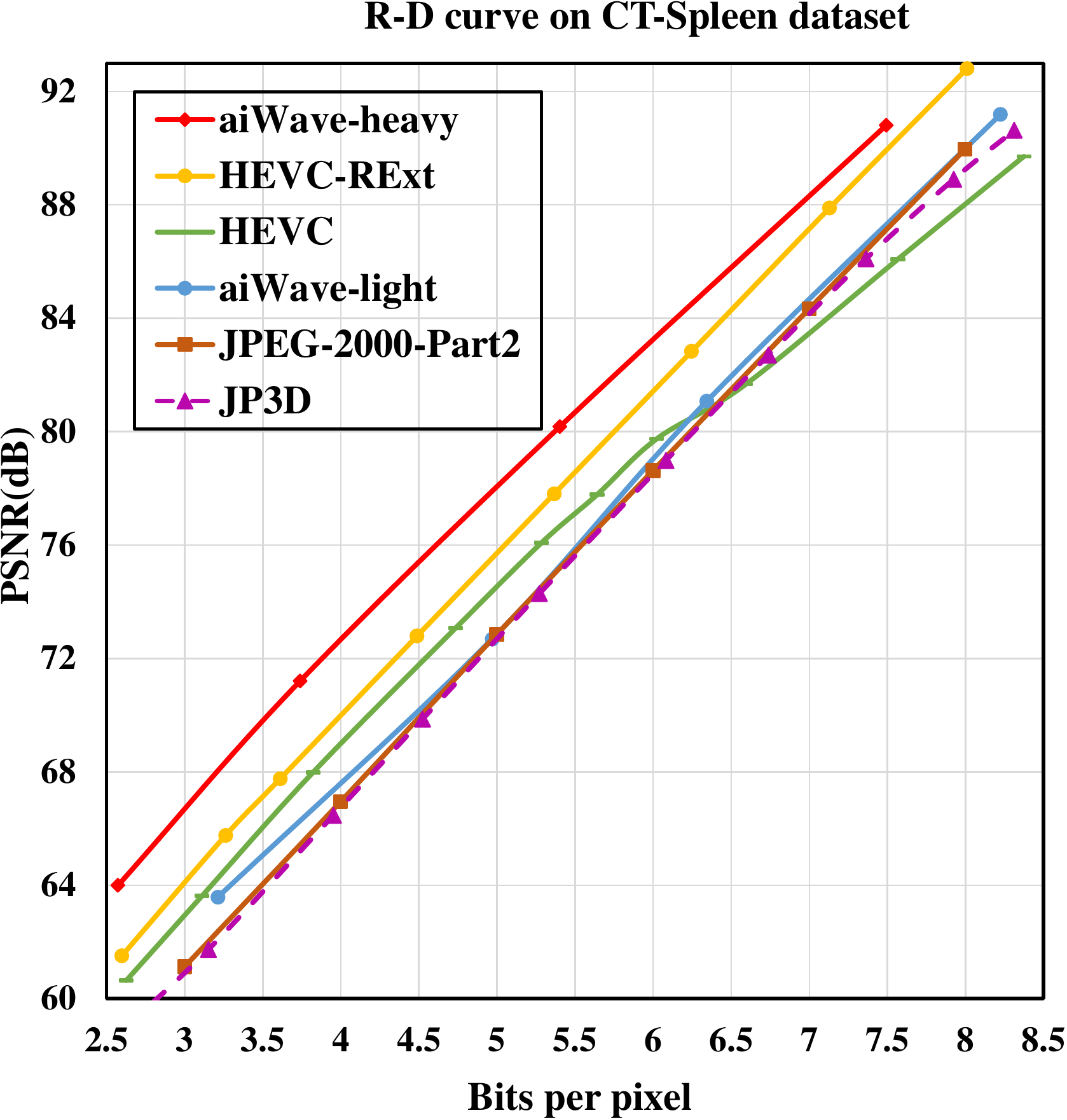}
    }
    
   \subfigure[MRI-heart dataset]
   {
   \includegraphics[width=0.62\columnwidth]{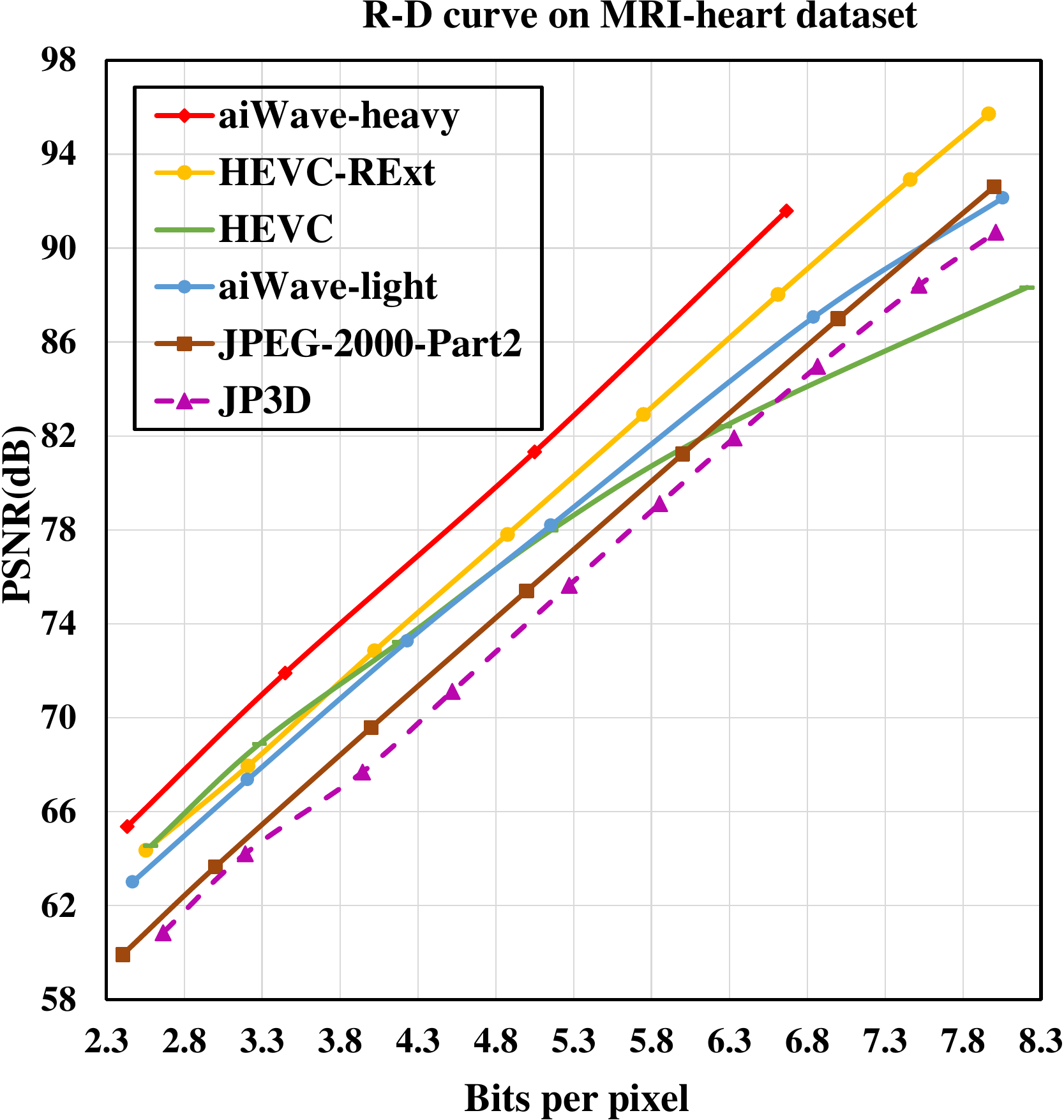}
   }
    \subfigure[Chaos\_CT dataset]
   {
   \includegraphics[width=0.62\columnwidth]{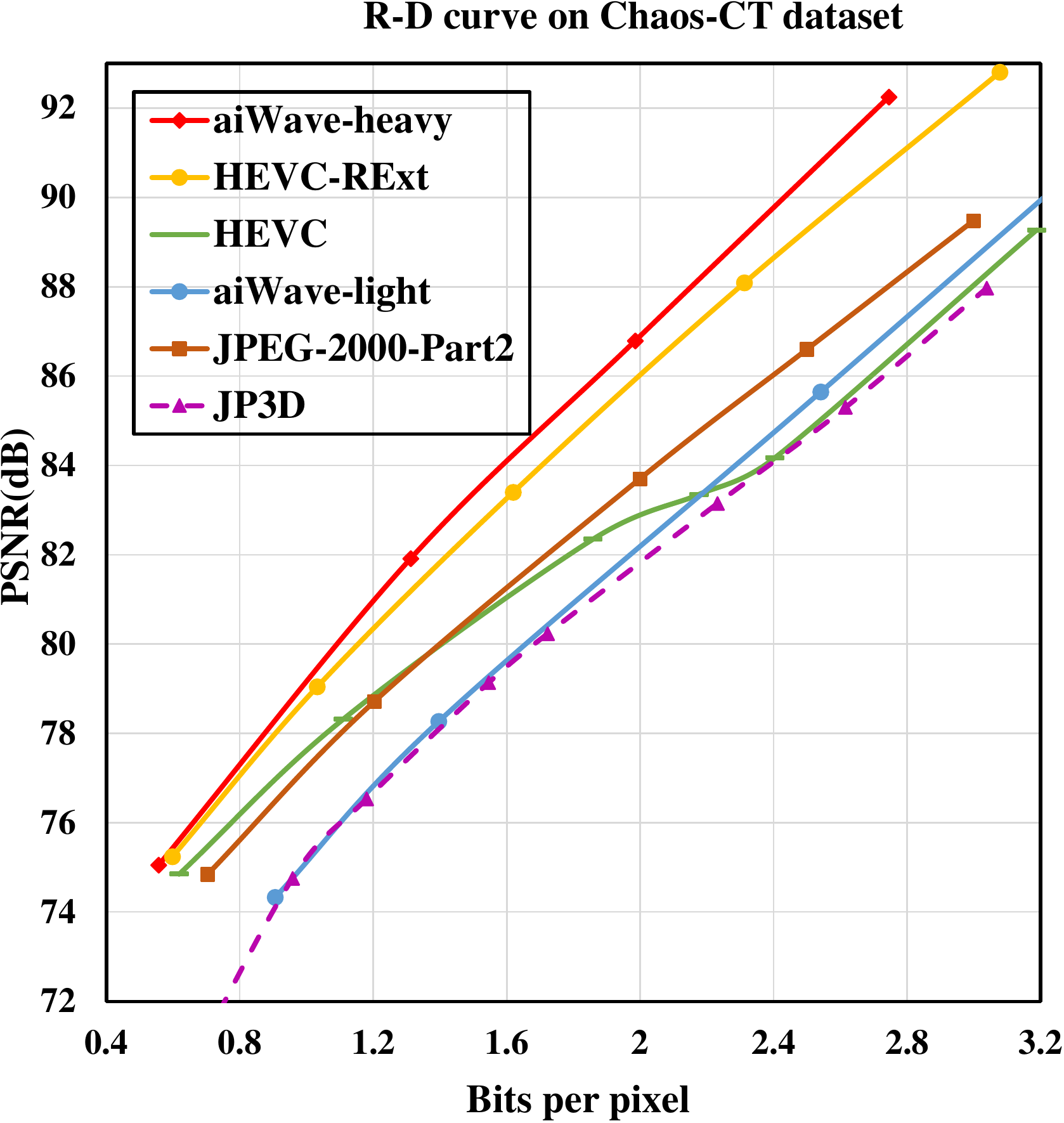}
   }
    \subfigure[Attention dataset]
   {
   \includegraphics[width=0.62\columnwidth]{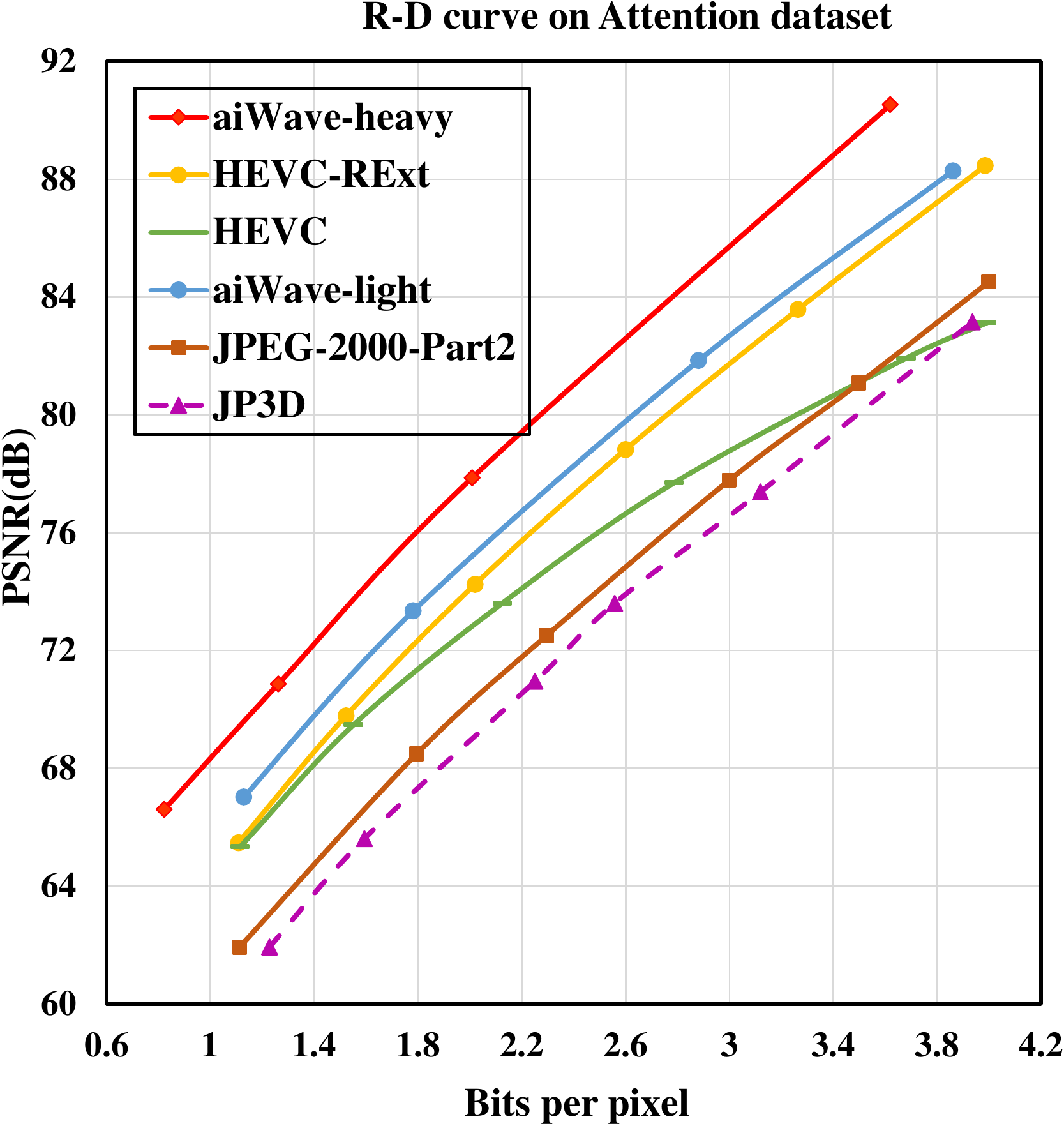}
   }

   \caption{R-D performance comparisons on FAFB, FIB-25, CT-Spleen, MRI-heart, Chaos-CT, and Attention datasets for lossy compression. "aiWave-heavy" stands for aiWave with context-based entropy coding module and "aiWave-light" stands for aiWave with factorized entropy coding module.}
   	\label{main}
    \centering
\end{figure*}
\subsection{Experiment settings}
We evaluate our aiWave on validation data with peak-signal-to-noise ratio (PSNR) and compare it with JP3D, JPEG-2000-Part2, HEVC, HEVC-RExt, and other learning-based methods. 

For JP3D, we use the software OpenJPEG 2.3.1\footnote{http://www.openjpeg.org/2019/04/02/OpenJPEG-2.3.1-released} with the default configuration. opj\_jp3d\_compress and opj\_jp3d\_decompress are employed for encoding and decoding, respectively. We specified the use of 3D wavelet transform and 3D code block coding. During the testing step, CDF 9/7 wavelet transform is used for lossy compression, and CDF 5/3 wavelet transform is used for lossless compression.

For JPEG-2000-Part2, the software Kakadu V6.1\footnote{https://kakadusoftware.com/} is used, and kdu\_compress and kdu\_expand are employed for encoding and decoding. The transform is set to DWT with 4 levels, and the kernel index is set to CDF 9/7 for lossy compression and CDF 5/3 for lossless compression.

For HEVC, we use the reference software HM-16.15\footnote{https://vcgit.hhi.fraunhofer.de/jvet/HM/-/tree/HM-16.15} with the random-access configuration. We set GOPSize to 16, IntraPeriod to -1, and InternalBitDepth to be consistent with the bit depth of each dataset, without RExt. 

For HEVC-RExt, the HM 16.15 software is used with the randomaccess-rext config file. We set GOPSize to 16, IntraPeriod to -1, and InternalBitDepth to be consistent with the bit depth of each dataset.

We first fix the transform and train the entropy coding and post-processing modules in our training process.
Then, we fix the entropy coding and post-processing modules and train the affine wavelet-like transform.
Finally, we train all modules end-to-end to obtain the final model.
The Adam algorithm with its default setting is adopted in the training process.
The learning rate is set to \(1e^{-4}\). 
The value of \(\lambda\) in \eqref{eq::optimization} is set to \{1, 4, 16, 64, 128\} to generate bitstream with variable bitrates.
The whole training process is carried out on a GTX 1080Ti GPU and takes about 6 days. 
For the CPU runtime, we used an Intel(R) Core(TM) i5-8265U CPU @ 1.60GHz, 1800Mhz, 4 cores, and 8 logical processors. The OS used is Microsoft Windows 10 Home Chinese Edition; the version number is 10.0.19042.
For the GPU runtime, we used the same GPU as that in the training process.

\begin{table*}
\centering
\caption{BD-PSNR / BD-rate improvements of different methods compared with JP3D}
\setlength{\tabcolsep}{5.5mm}{
\begin{tabular}{cccccc}
\toprule[0.8pt]
Datasets & JPEG-2000-Part2 & HEVC & HEVC-RExt & aiWave-light  &aiWave-heavy\\ 
\toprule[0.8pt]
FAFB & -0.243 / 1.300\% & 2.662 / -16.641\% & 2.606 / -16.058\% & 1.766 / -10.639\% & 3.817 / -23.817\%  \\ 
FIB-25 &1.567 / -12.401\% &3.369 / -25.653\% & 3.360 / -25.974\% & 1.882 / -14.157\% & 4.091 / -30.332\%  \\ 
CT-Spleen &0.232 / -0.753\% &1.335 / -5.015\% & 3.130 / -11.652\% & 0.507 / -2.231\% &5.392 / -20.817\%  \\ 
MRI-heart &1.266 / -4.606\% &2.604 / -11.200\% & 4.430 / -17.234\% & 3.073 / -12.545\% & 6.878 / -28.511\%  \\ 
Chaos-CT &2.061 / -20.458\% &1.953 / -16.289\% & 4.124 / -45.158\% &0.332 / -3.073\% & 4.886 / -55.594\%  \\ 
Attention &1.103 / -6.470\% &3.069 / -19.853\% & 4.991 / -32.153\% & 6.047 /-40.480\% & 8.885 / -67.805\%  \\ 
\bottomrule[0.8pt]
\end{tabular} }
\label{BD-PSNR}
\end{table*}

\begin{figure}
	\centering
		\includegraphics[width=0.80\columnwidth]{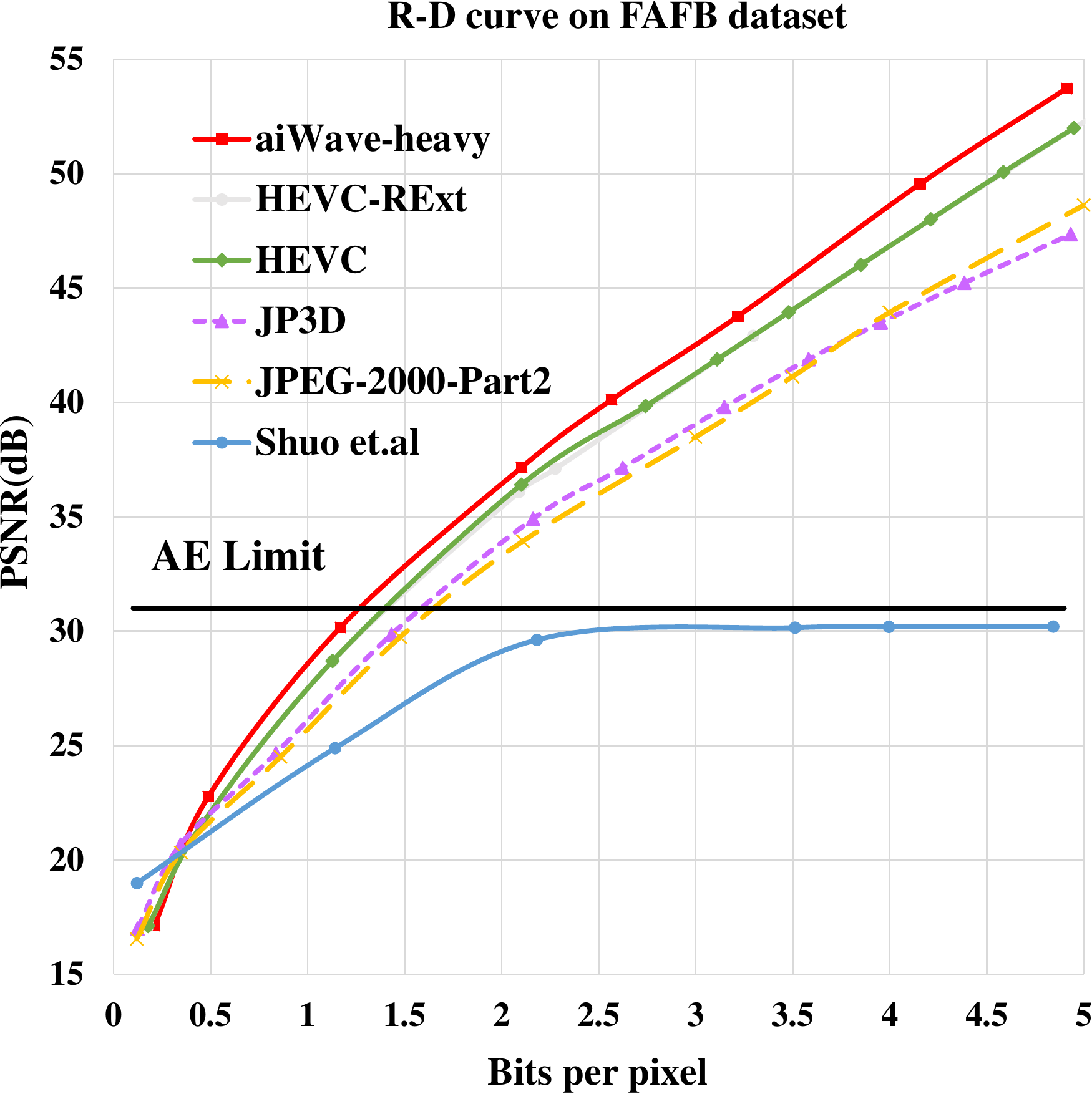}
    	\caption{R-D performance for aiWave-heavy when compared with autoencoder-based method on FAFB dataset.}
    	\label{shuo}
	\centering
\end{figure}

\subsection{Lossy compression performance}
\subsubsection{Comparison with traditional compression methods}
Fig.~\ref{main} shows the rate-distortion (R-D) curve comparison between the proposed aiWave and the traditional compression methods in lossy case, where the rate and distortion are measured by bits per pixel (bpp) and PSNR, respectively.
Our aiWave-heavy outperforms all traditional compression methods for all bitrates and even surpasses HEVC-RExt by a large margin. 
The phenomenon that aiWave-heavy leads to a much better performance than aiWave-light indicates that the context-based entropy coding module further removes redundancies compared with the factorized entropy coding module. 
For the aiWave-light, it outperforms JP3D due to the use of the affine wavelet-like transform and post-processing modules. 
In order to quantitatively measure the performance, the average Bjontegaard Delta-PSNR (BD-PSNR) and Bjontegaard Delta-Rate (BD-Rate) improvements of aiWave-heavy compared with the other methods are shown in Table~\ref{BD-PSNR}.
Taking the FIB-25 dataset as an example, ``$4.091$dB / $-30.332$\%'' means that aiWave-heavy improves $4.091$dB compared with JP3D, which is equivalent to saving about \(-30.332\)\% of the bit rate under the same quality.
In addition, aiWave-light also outperforms JP3D by \(1.882\)dB.

\subsubsection{Comparison with learning-based methods}
Fig.~\ref{shuo} shows the comparison between aiWave-heavy and the current state-of-the-art volumetric image compression method based on autoencoder\cite{gao2020volumetric}. 
The experimental results show that the method of Gao \emph{et al.} only achieves better performances at very low bitrates.  
As the bitrate increases, their method converges, and the PSNR does not increase. 
This phenomenon is the same as ``AE Limit" in 2-D image compression, as claimed in \cite{helminger2020lossy}. 
The autoencoder is irreversible and thus cannot achieve high-quality reconstruction. 
As a consequence, this experiment further illustrates that a reversible and efficient transform like affine wavelet-like transform is more suitable for volumetric image compression.

\subsection{Lossless compression performance}
\subsubsection{Comparison with traditional compression methods}
Table~\ref{lossless-traditional} shows the average bits per pixel of various algorithms on the test datasets. 
It can be seen that the average bpp of aiWave-heavy is the smallest to achieve lossless compression. 
In addition, aiWave-light also performs better than JP3D on all datasets. 
The bpp reduction of aiWave on the Attention dataset is the most significant among them. 
Compared with JP3D and JPEG-2000-Part2, the bits per pixel is reduced by $36.8\%$ and $40.9\%$ when using aiWave-heavy.
Compared with HEVC and HEVC-RExt, the bits per pixel are reduced by $39.4\%$ and $39.4\%$ when using aiWave-heavy.
Meanwhile, aiWave-light also saves 18.4\% bits on average compared with JP3D. 

\subsubsection{Comparison with learning-based methods}
We compared aiWave with the current state-of-the-art lossless compression methods ICEC\cite{chen2022exploiting} and L3C\cite{mentzer2019practical}. 
ICEC is a lossless compression network for 3D images designed to exploit the correlation within intra and inter frames. 
At the same time, L3C is another representative method for end-to-end lossless compression. 
We select the MRNet dataset used in their paper \cite{chen2022exploiting} and conduct experiments in the same setting as theirs. 
The experiment results are shown in Table \ref{lossless-learning}. 
All the anchors are directly copied from their paper except for our method. 
Experiment results show that our aiWave-heavy can outperform all methods on three datasets. 
For MRNet-axial, MRNet-coronal, and MRNet-sagittal, aiWave saves about 2.1\%, 1.0\%, and 2.9\% in bits per pixel, respectively. 
In addition to these two state-of-the-art lossless compression methods, we also found that aiWave-heavy outperforms other traditional codecs mentioned in their paper, including PNG, JPEG-LS, JPEG-2000, HEVC-Intra, HEVC-RExt-Intra, and FFV1.

\subsection{Runtime Analysis}
Table~\ref{codec time} shows the complexities of the proposed framework in terms of encoding/decoding time and the number of parameters.  
Note that the encoding/decoding time when using GPU is obtained through multi-process parallel acceleration. 

It can be seen that the encoding time of aiWave-light is shorter than that of JP3D because our factorized entropy coding module only relies on the current coefficients when encoding and decoding and thus can be accelerated in parallel. 
When decoding, due to the post-processing module, the decoding time of aiWave is slightly slower than JP3D. 
In other words, our aiWave-light and JP3D have a comparable runtime, but the compression performance of aiWave-light is much better. 
Our aiWave-heavy has a slower speed and a larger amount of parameters compared with aiWave-light. 
This is because each pixel to be encoded depends on all the encoded pixels in the context-based entropy coding module. 
It trades larger runtime for a significant performance improvement. 
We also notice that since the the traditional codec has been optimized and accelerated many times, the decoding speed of HEVC is very fast. 
Moreover, Kakadu software is known as a high-speed implementation of JPEG-2000, and a large number of parallelization is designed to save time. 
To this end, Kakadu has the fastest encoding speed. In the decoding stage, since Kakadu can only get the decode images of one channel at a time, its decoding time is longer than that of JP3D, HEVC and HEVC-RExt.

\begin{table}
\centering
\caption{Average bits per pixel on six datasets for lossless compression}
\setlength{\tabcolsep}{4.5mm}{
\begin{tabular}{cccc}
\toprule[0.8pt]
Methods &Axial &Coronal&Sagittal\  \\
\toprule[0.8pt]
PNG\cite{roelofs1999png} & 5.357 & 4.577 & 5.582 \\
JPEG-LS\cite{weinberger2000loco} & 4.911& 4.073& 5.239 \\
JPEG-2000\cite{taubman2002jpeg2000} & 5.016 & 4.170 & 5.309 \\
JP3D\cite{bruylants2009jp3d} & 4.977 & 4.148 & 5.278\\
JPEG-2000-Part2\cite{boliek2000information} & 5.001 & 4.155 & 5.296\\
HEVC-Intra\cite{sullivan2012overview} &5.309 & 4.470 & 5.639\\
HEVC\cite{sullivan2012overview} &5.190 & 4.467 & 5.579\\
HEVC-RExt-Intra\cite{parikh2017high}  & 5.235 & 4.346 & 5.632\\
HEVC-RExt\cite{parikh2017high} & 5.110 & 4.204 & 5.502 \\ 
FFV1\cite{niedermayer2013ffv1} & 4.859 & 4.028 & 5.185\\
L3C\cite{mentzer2019practical} & 5.156 & 4.446 & 5.516\\
ICEC\cite{chen2022exploiting} & 4.642 & 3.841 &4.974\\
aiWave-heavy  & \textbf{4.545} & \textbf{3.804} &\textbf{4.829}\\
\bottomrule[0.8pt]
\end{tabular}}
\label{lossless-learning}
\end{table}

\begin{table*}
\centering\caption{Average bits per pixel on six datasets for lossless compression}
\setlength{\tabcolsep}{6mm}{
\begin{tabular}{lcccccc}
\toprule[0.8pt]
Datasets &JP3D & JPEG-2000-Part2 & HEVC & HEVC-RExt & aiWave-light  & aiWave-heavy \\ 
\toprule[0.8pt]
FAFB\cite{zheng2018complete} &6.406 &6.541 & 6.260 & 6.260 & 5.940 &{\textbf{5.636}} \\ 
FIB-25 & 5.660 &5.025 & 5.171  & 5.171 & 4.723 & {\textbf{4.608}}   \\
CT\cite{simpson2015chemotherapy} & 10.765 & 11.374 & 10.689 & 10.704 & 10.557 & {\textbf{9.789}}  \\ 
MRI\cite{tobon2015benchmark} & 10.394 &11.581 & 10.551 & 10.569 & 10.392 & {\textbf{9.161}}   \\ 
Chaos-CT\cite{simpson2015chemotherapy} & 5.997 &6.059 & 5.975 & 5.975 & 5.972 & {\textbf{5.077}}  \\ 
Attention\cite{tobon2015benchmark} & 7.436 &7.964 & 7.756  & 7.758 & 7.321 & {\textbf{4.703}}   \\ 
\bottomrule[0.8pt]
\end{tabular} }
\label{lossless-traditional}
\end{table*}

\begin{table*}
\centering
\caption{Average running time per image on the FAFB dataset (in seconds) and the number of training parameters}
\fontsize{8.3}{9}\selectfont
\setlength{\tabcolsep}{4.5mm}{
\begin{tabular}{lcccccccc}
\toprule[0.8pt]
~&JP3D & JPEG-2000-Part2 & HEVC & HEVC-RExt & \multicolumn{2}{c}{aiWave-light} & \multicolumn{2}{c}{aiWave-heavy} \\
~&~&~&~&~&CPU&GPU& CPU&GPU \\ 
\toprule[0.6pt]
Avg. Enc. Time&0.487&\textbf{0.066}&9.617&12.647&5.528&0.281&986.966&943.211 \\
Avg. Dec. Time&0.243&0.576&\textbf{0.052}&{0.067}&5.947&0.317&982.032&956.08\\
Parameters&-&-&-&-&\multicolumn{2}{c}{\textbf{21.3MB}}& \multicolumn{2}{c}{695.5MB} \\
\bottomrule[0.8pt]
\end{tabular}}
\label{codec time}
\end{table*}
\subsection{Ablation study}
\subsubsection{Comparison of affine wavelet-like transform, additive wavelet-like transform and traditional CDF 9/7 wavelet transform}
\begin{figure}
    \centering
	\subfigure[R-D curve on FAFB dataset]
    {\includegraphics[width=0.46\columnwidth]{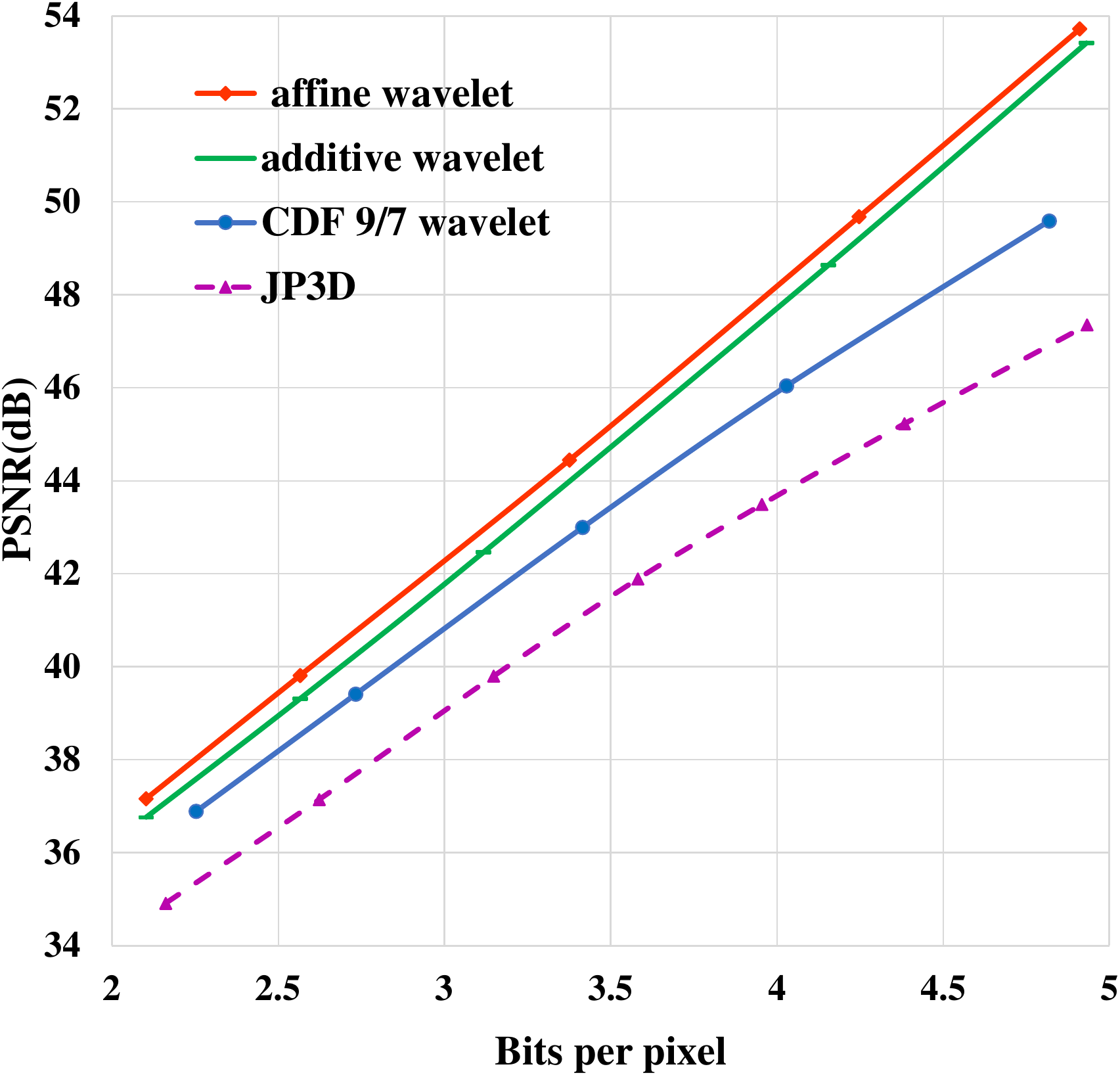}
     \label{a1}}
 	\subfigure[R-D curve on MRI dataset]
    {\includegraphics[width=0.45\columnwidth]{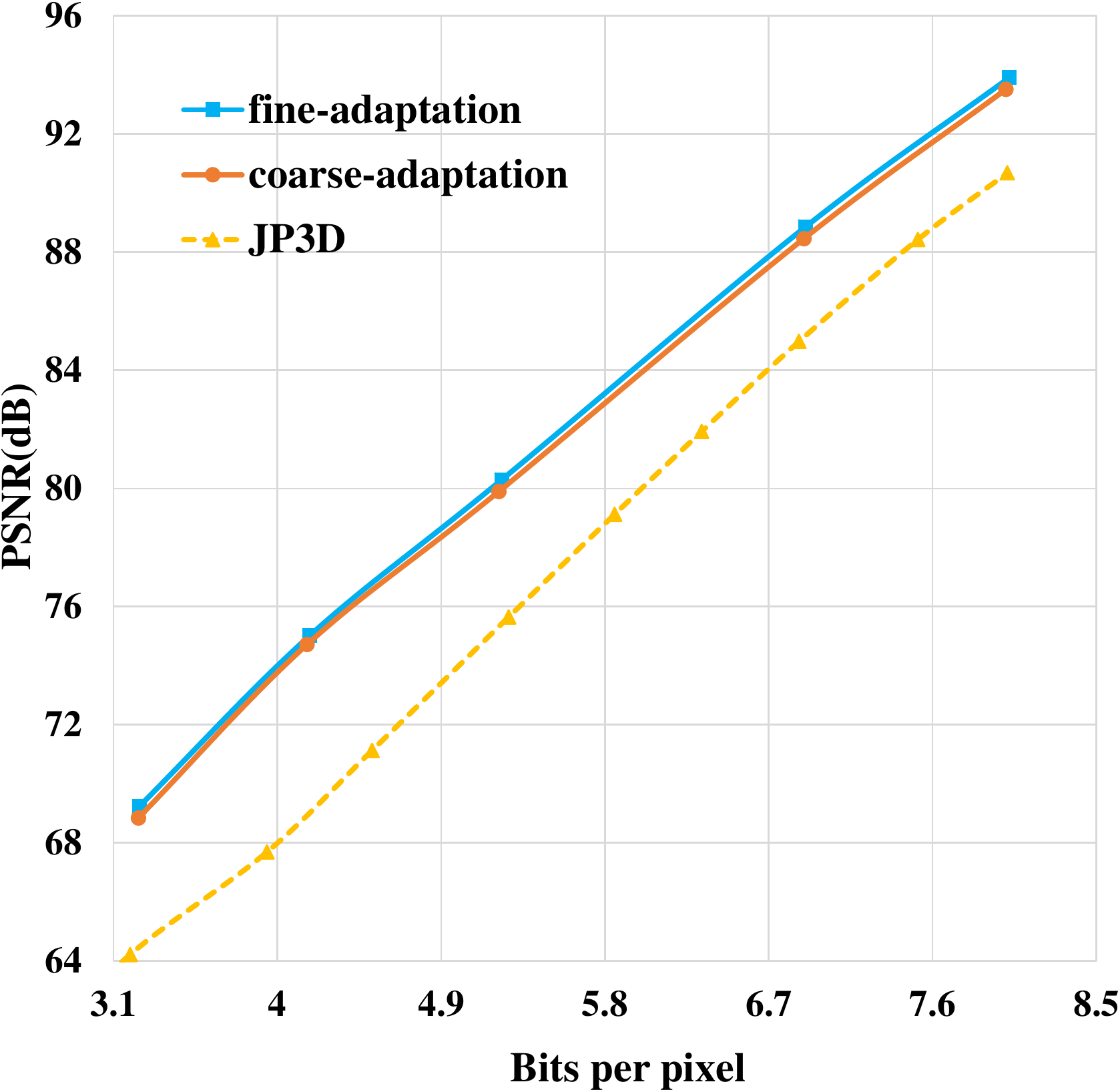}
     \label{b1}}
    \caption{Subfigure (a) compares the affine wavelet-like transform, the additive wavelet-like transform, the traditional CDF 9/7 wavelet and JP3D on FAFB dataset in terms of R-D curve. Subfigure (b) shows the comparison between fine-adaptation affine wavelet-like transform, coarse-adaptation affine wavelet-like transform and JP3D on MRI-heart dataset in terms of R-D curve.}
    \label{ablation_results}
   \centering
\end{figure}

Fig.~\ref{a1} shows the performance comparisons among affine wavelet-like transform, additive wavelet-like transform, and traditional CDF 9/7 wavelet transform. 
The BD-rate improvements compared with JP3D in Fig.~\ref{a1} are shown in Table~\ref{BD-Rate-affine}.
The additive wavelet-like transform is obtained by removing the affine map in the affine wavelet-like transform.
We can see from Fig.~\ref{a1} that our affine wavelet-like transform significantly outperforms traditional CDF 9/7 wavelet transform by $1.859$dB on average.
In addition, our affine wavelet-like transform performs better than the additive wavelet-like transform on the FAFB dataset, which demonstrates that predicting a context-adaptive affine map indeed brings some performance improvements. 
Furthermore, the performance of our additive wavelet-like transform is much higher than that of traditional wavelets, which demonstrates that our 3-D learnable wavelet transform significantly improves the performance by replacing traditional manual design through training with a large number of samples.

\subsubsection{Comparison of coarse-adaptation and fine-adaptation affine wavelet-like transform}
\label{fine-coarse}
Fig.~\ref{b1} shows the comparison between the coarse-adaptation and fine-adaptation affine wavelet-like transforms.
The BD-rate improvements compared with JP3D in Fig.~\ref{b1} are shown in Table~\ref{BD-adaptation}.
As we described in Section \ref{affine section}, the coarse-adaptation affine wavelet-like transform is obtained through sharing parameters for all spatial positions.
The experimental results show that the coarse-adaptation affine wavelet-like transform leads to a bit of performance degradation. 
However, the coarse-adaptation version has fewer parameters. We also observe that coarse-adaptation version converges faster, and can be trained more stable during training.

\subsubsection{Comparison of different weight sharing strategies for different volumetric images.}
\label{weight results}
\begin{figure}
	\subfigure[Anisotropy dataset]
	{\includegraphics[width=0.45\columnwidth]{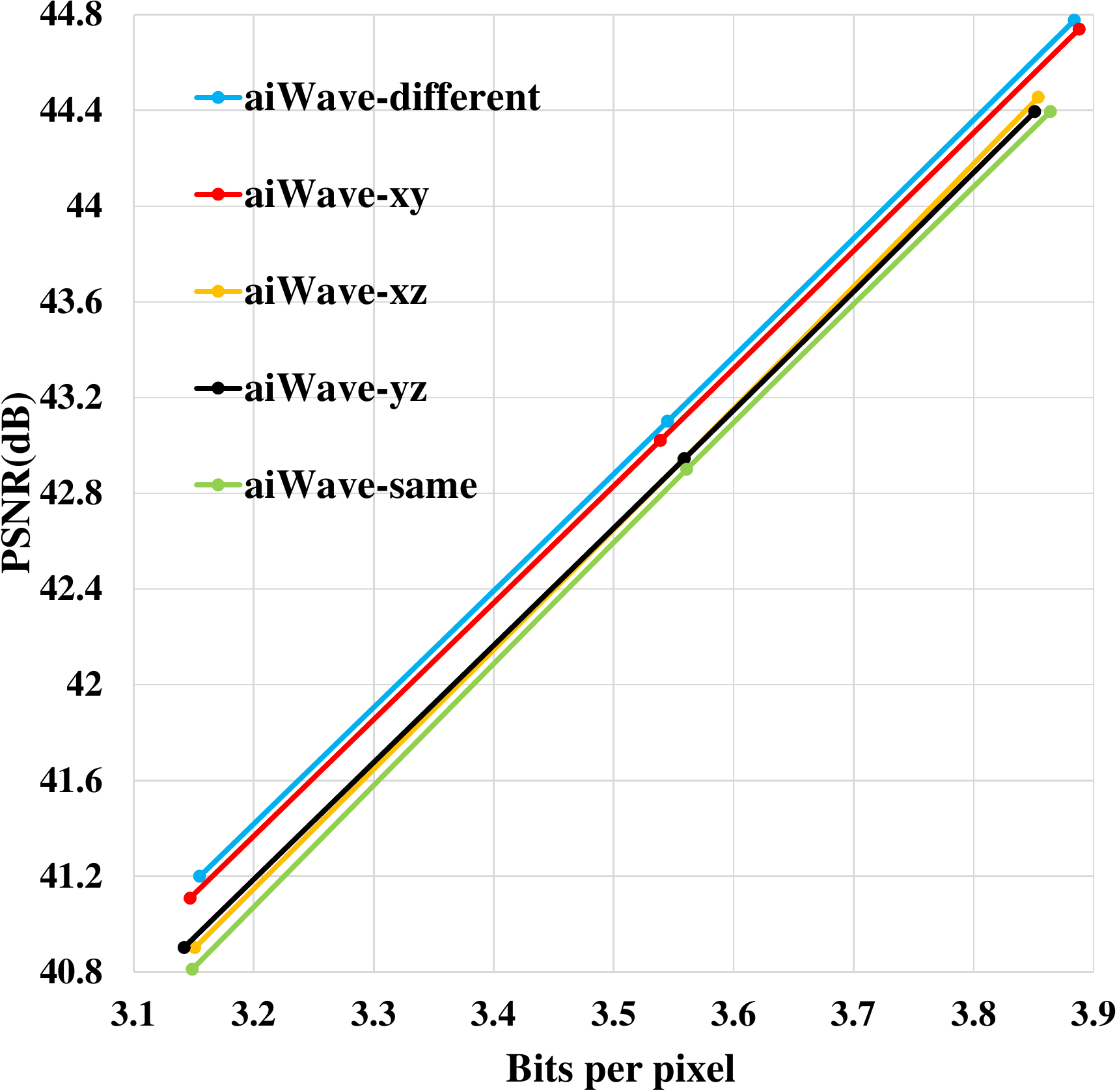}}
    \subfigure[Isotropy dataset]
	{\includegraphics[width=0.47\columnwidth]{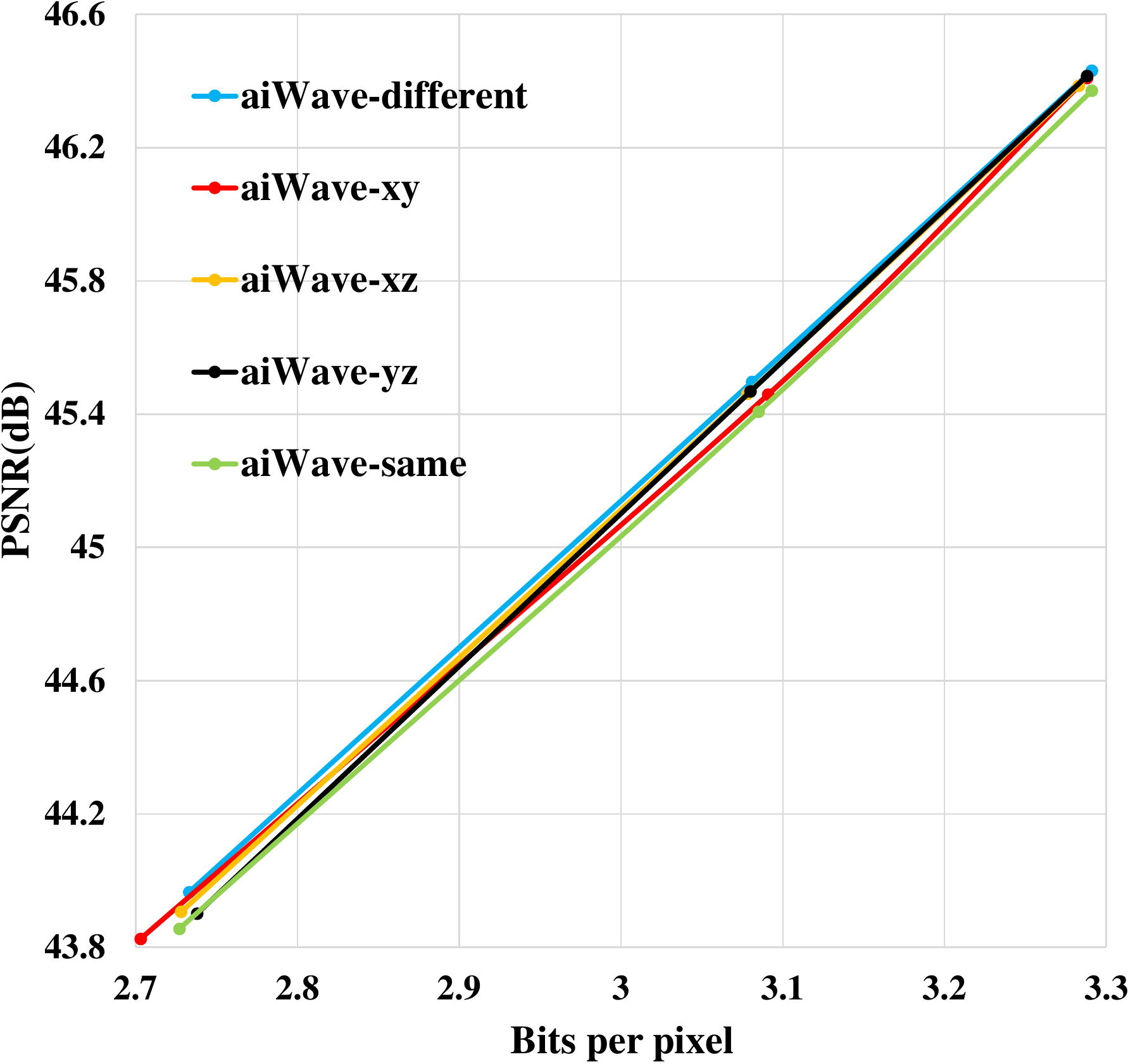}}
    \caption{Comparisons of different parameter sharing strategies. aiWave-different represents that the update and prediction networks in the three directions do not share parameters. aiWave-same shows that all three directions share parameters. aiWave-xy or xz or yz means sharing parameters only in the x and y or x and z or y and z directions. }
    \label{weight_results}
\end{figure}

Fig.~\ref{weight_results} shows the performance of different weight sharing strategies for affine wavelet-like transform mentioned in Section~\ref{weight}.
The BD-rate improvements in Fig.~\ref{weight_results} are shown in Table \ref{BD-PSNR-anisotropy}.
Anisotropic FAFB dataset and isotropic FIB-25 dataset are applied to represent two kinds of volumetric images.  

For an anisotropic dataset, we find that sharing parameters in the \(x\) and \(y\) directions while not in the \(z\) direction lead to almost the same performance as not sharing parameters in all three directions.
Note that using independent parameters in the \(z\) direction is about 0.5dB better than using independent parameters in the \(x\) or \(y\) direction alone, even if they have the same number of parameters. 
This phenomenon indicates that the characteristics of the \(z\)-direction of anisotropic datasets are different from those of \(x\) and \(y\) directions.
Therefore, we can use independent parameters for the \(z\) direction.

For isotropic datasets, we find that even if we do not share parameters in three directions, the performance does not improve significantly compared with sharing parameters in all three directions. 
That further demonstrates that the isotropic dataset has consistent properties in all three directions. 
Therefore, all parameters can be shared to reduce complexities without affecting the compression performance.


\begin{table}
\centering
\caption{BD-PSNR and BD-rate improvements of different wavelets compared with JP3D}
\setlength{\tabcolsep}{2mm}{
\begin{tabular}{cccccc}
\toprule[0.8pt]
Datasets & CDF 9/7 wavelet &additive wavelet  & affine wavelet \\ 
\toprule[0.8pt]
FAFB  & 1.958 / -12.865\% & 3.356 / -20.308\% &  3.187 / -23.817\%   \\ 
\bottomrule[0.8pt]
\end{tabular} }
\label{BD-Rate-affine}
\end{table}

\begin{table}
\centering
\caption{BD-PSNR and BD-rate improvements of coarse-adaptation and fine-adaptation affine wavelet-like transform compared with JP3D}
\setlength{\tabcolsep}{2mm}{
\begin{tabular}{cccccc}
\toprule[0.8pt]
Datasets &coarse-adaptation &fine-adaptation \\ 
\toprule[0.8pt]
MRI-heart & 4.471/ -16.481\% & 4.772/ -17.644\%   \\ 
\bottomrule[0.8pt]
\end{tabular} }
\label{BD-adaptation}
\end{table}

\begin{table}
\centering
\caption{BD-PSNR and BD-rate improvements of different strategies compared with the aiWave-same}
\setlength{\tabcolsep}{0.7mm}{
\begin{tabular}{cccccc}
\toprule[0.8pt]
Datasets  & aiWave-xz & aiWave-yz  & aiWave-xy&aiWave-different \\ 
\toprule[0.8pt]
FAFB  & 0.065 / -0.374\% &  0.071 / -0.411\% & 0.246 / -1.416\%&0.297 / -1.175\%  \\ 
FIB-25 & 0.075 / -0.552\% & 0.063 / -0.446\% &  0.034 / -0.288\% &0.101 / -0.749\%  \\ 
\bottomrule[0.8pt]
\end{tabular} }
\label{BD-PSNR-anisotropy}
\end{table}

\subsubsection{Energy compaction analysis}
\begin{figure}
    \centering
	\subfigure[]
    {\includegraphics[width=0.48\columnwidth]{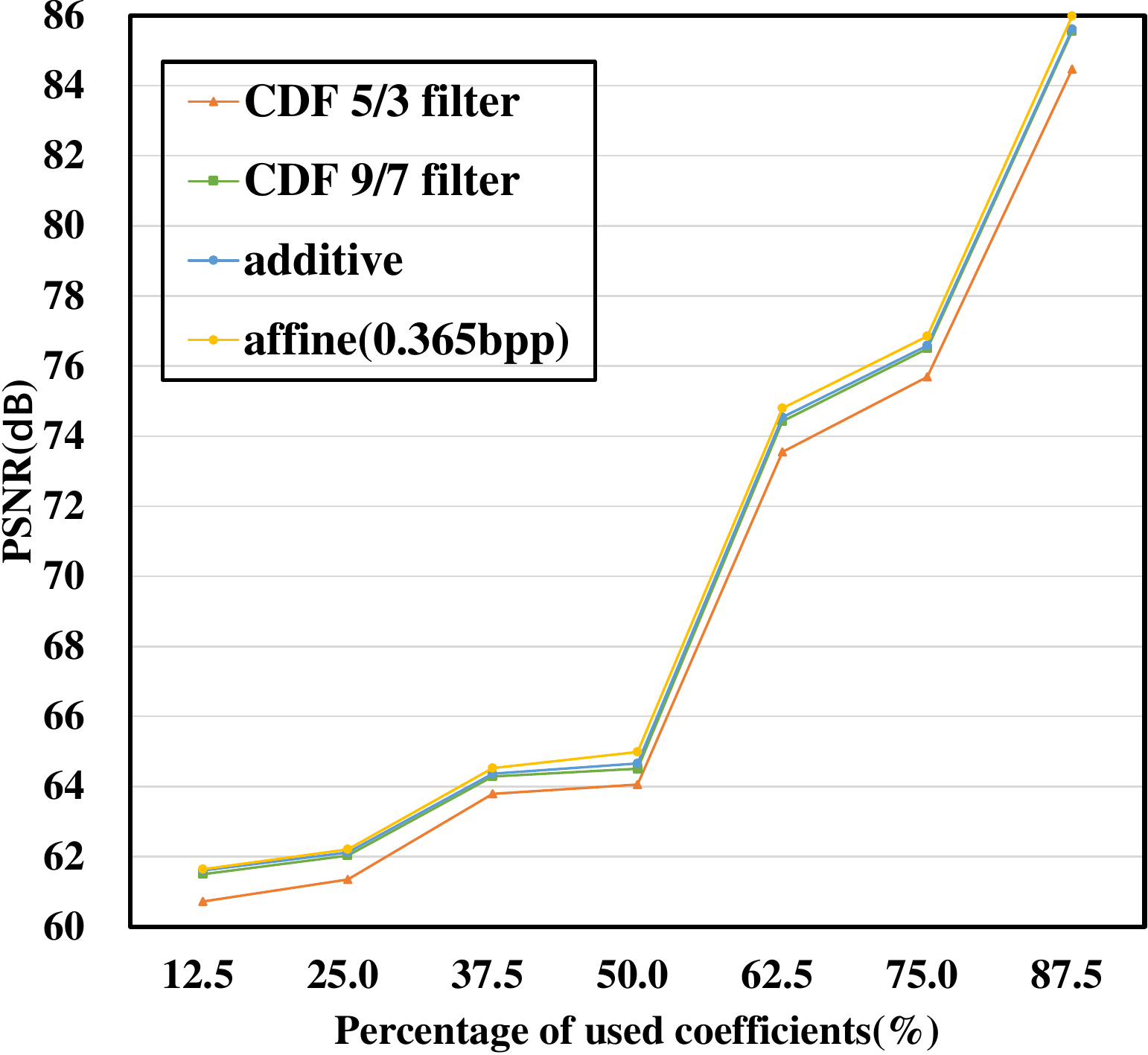}
     \label{a}}
 	\subfigure[]
    {\includegraphics[width=0.48\columnwidth]{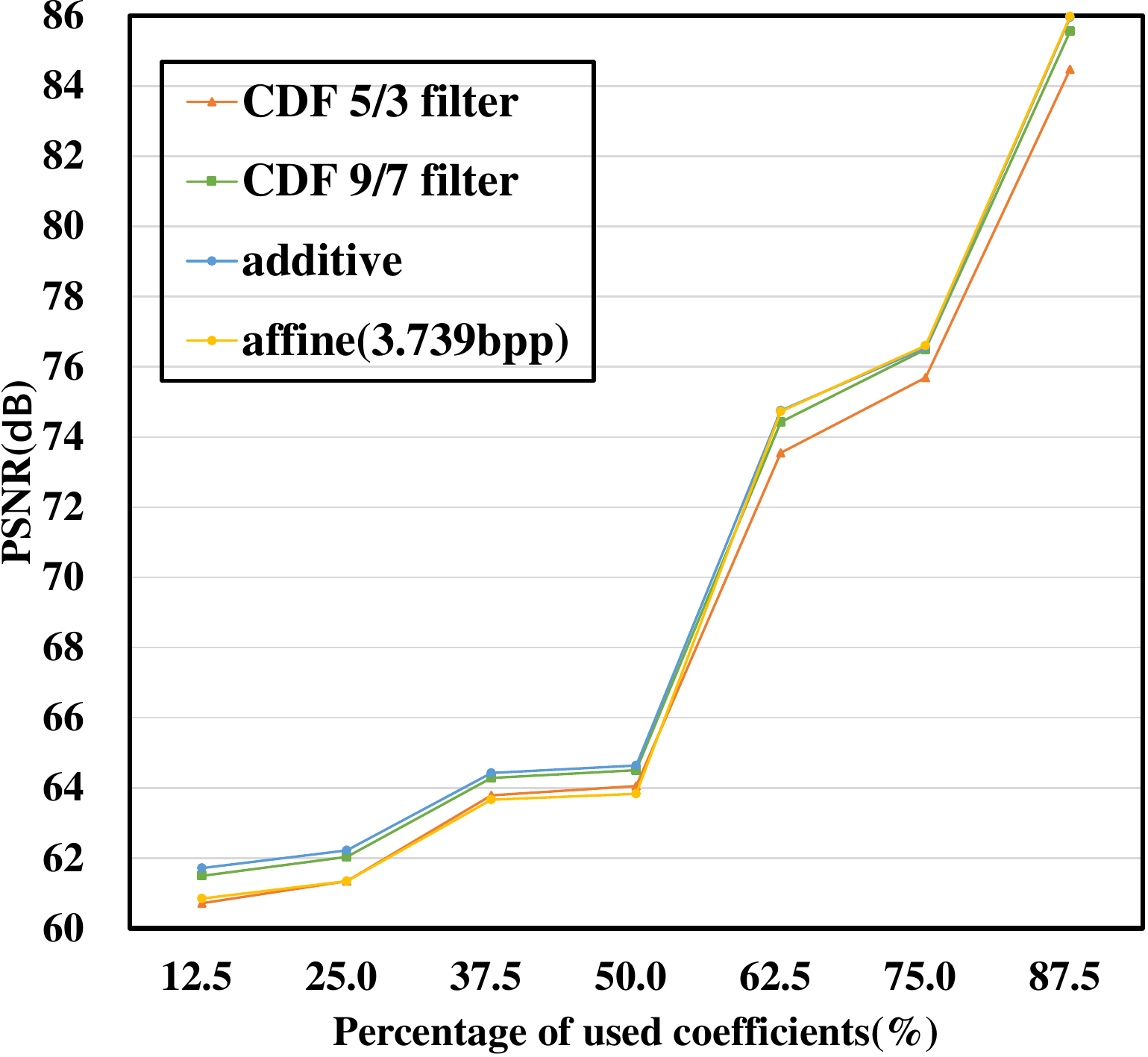}
     \label{b}}
    \caption{Comparison of CDF 5/3 wavelet transform, CDF 9/7 wavelet transform, affine wavelet-like transform and additive wavelet-like transform in terms of reconstructed PSNR of partial subband coefficients on the Spleen-CT dataset}
    \label{visual}
   \centering
\end{figure}

To quantitatively demonstrate the energy concentration of aiWave, we compared the CDF 5/3 wavelet, CDF 9/7 wavelet, affine wavelet-like transform, and additive wavelet-like-transform (The additive wavelet-like transform is obtained by removing the affine map in the affine wavelet-like transform) on the Spleen-CT dataset with the reconstructed PSNR of partial subband coefficients. The experiment results in Fig. \ref{visual} and Table \ref{BD-energy} show that the additive wavelet-like transform has a better energy concentration than the CDF 9/7 wavelet and  CDF 5/3 wavelet at all bit rates. The affine wavelet-like transform has the best energy concentration at a low bit rate while does not have a good energy concentration at high bitrates. This is because all prediction and update networks of the additive wavelet-like transform share parameters during training, limiting it to make each subband's energy concentration better to achieve a better performance. However, all prediction and update networks of affine wavelet-like transform do not share parameters during training, and different subbands do not need to become energy concentrated to achieve a good performance. Especially when the bit rate is relatively high (right picture in Fig. \ref{visual}), almost all low-frequency subbands need to be encoded to get a high reconstruction quality. So the affine wavelet-like transform only guarantees the energies of the high-frequency subbands to be low.

\subsection{Limitations of the proposed framework}
The proposed framework achieves significant BD-rate savings compared with the traditional methods by training data-dependent wavelets.
However, it seems that we have not made full use of the characteristics of ``data-dependent'' as we expected an even larger performance improvement.
First, we consider that 3D biomedical images have their common statistical characteristics such as high spatial correlations and many low-frequency components. 
The data-independent transform such as CDF 9/7 wavelet can characterize these common features effectively. 
The only thing the data-dependent transform can do is to characterize the unique characteristic belonging to each image.
Second, our training is minimizing the average rate-distortion loss for all the images in a dataset. 
We have not performed online optimization for each specific image, which can be quite complex, to achieve a better balance between the complexity and the performance.

\section{Conclusion}
This paper proposes an end-to-end volumetric image compression framework aiWave, which supports both lossy and lossless compression.
We design a trained affine wavelet-like transform to obtain more compact representations of images.
Then, we introduce a trainable affine map to enable a content-related wavelet-like basis in different spatial regions of images. 
After that, we introduce the weight sharing strategies according to the volumetric data characteristics in the axial direction to reduce the number of parameters. 
Finally, different entropy coding modules are explored to handle different application scenarios. 
The experimental results demonstrate the superiority of our proposed aiWave compared with the state-of-the-art methods on a variety of 3-D image datasets.

\begin{table}
\centering
\caption{Average reconstruction PSNR improvement of CDF 9/7 wavelet, affine wavelet-like transform and additive wavelet-like transform compared with CDF 5/3 wavelet}
\setlength{\tabcolsep}{2mm}{
\begin{tabular}{cccccc}
\toprule[0.8pt]
bpp &CDF 9/7 wavelet &additive  &affine 
\\ 
\toprule[0.8pt]
0.365 & 0.685 & 0.780 &0.943 \\ 
3.739 & 0.685 & 0.883 &0.209  \\ 
\bottomrule[0.8pt]
\end{tabular} }
\label{BD-energy}
\end{table}

 \bibliographystyle{ieeetr}
 \bibliography{refer}

\end{document}